\RequirePackage{amsmath}
\documentclass[12pt]{iopart}
\usepackage{iopams} 
\usepackage{graphicx} \graphicspath{{figs/}}

\begin{document}

\title{Geometrically thick tori around compact objects with a quadrupole moment}

\author{Jan-Menno Memmen and Volker Perlick}

\address{ZARM, University of Bremen, 28359 Bremen, Germany. 
\\
Email:  jan.memmen@zarm.uni-bremen.de, volker.perlick@zarm.uni-bremen.de}

\begin{abstract}
We study geometrically thick perfect-fluid tori with constant specific angular momentum, 
so-called ``Polish doughnuts", orbiting deformed compact objects with a quadrupole 
moment. More specifically, we consider two different asymptotically flat, static and 
axisymmetric vacuum solutions to Einstein's field equation with a non-zero quadrupole 
moment, the $q$-metric and the Erez-Rosen spacetime. It is our main goal to find features 
of Polish doughnuts in these two spacetimes which qualitatively distinguish them from 
Polish doughnuts in the Schwarzschild spacetime. As a main result we find that, for both
metrics, there is a range of positive (Geroch-Hansen) quadrupole moments which
allows for the existence of double tori. If these double tori fill their Roche
lobes completely, their meridional cross-section has the shape of a fish, with
the body of the fish corresponding to the outer torus and the fish-tail corresponding
to the inner torus. Such double tori do not exist in the Schwarzschild spacetime. 
\end{abstract}

%
%
%
%
%

\section{Introduction}
As no signal can escape from a black hole, detecting a black hole is a non-trivial task.
Recently, gravitational waves from merging black
holes have been observed \cite{LIGO2016}. However, these events are still comparatively rare, 
and they require two black holes spiralling onto each other. Static
or stationary black holes cannot be detected this way, as they do not emit 
any gravitational waves. Hence, the only way of observing such a black hole is 
via its influence on matter or light rays in its vicinity.

For studying the gravitational influence on matter one may consider a black hole 
that is surrounded by a geometrically thick perfect-fluid torus. An analytical
treatment is possible if one assumes that (i) the matter can be approximated as 
stationarily rotating (over a comparatively long time), without actually approaching 
the black hole, and (ii) the self-gravity of the fluid body can be neglected. For 
the simplest cases of a Schwarzschild and a Kerr black hole, such stationarily 
rotating perfect-fluid tori have been found, in the form of exact analytical solutions 
to the relativistic Euler equation, by a group of Polish scientists 
(see Jaroszy{\'n}ski et al. \cite{bib:donut-ss-kerr} for the original article, and also 
Rezzolla and Zanotti \cite{bib:rel-hydro} for a review), and they are now known as 
\emph{Polish doughnuts}. 

Such tori exist not only around black holes, but also around any other
object that is sufficiently compact. As such objects could be easily mistaken 
for black holes, they are known as \emph{black hole impostors} or \emph{black hole
mimickers}. Recall that according to the black-hole uniqueness theorem a black 
hole is spherically symmetric if it is non-rotating and isolated. More
specifically, this theorem implies that, if the spacetime around a black hole 
is a static and asymptotically flat solution to Einstein's vacuum field equation, 
then it is the Schwarzschild solution, i.e., all of its (Geroch-Hansen) multipole 
moments vanish with the exception of the monopole term. By contrast, the 
asymptotically flat and static vacuum spacetime around a black-hole impostor may 
have arbitrary mass multipole moments. In particular, it may have a non-zero 
quadrupole moment which, for a non-rotating and axisymmetric object, could be 
considered as the leading-order deviation from the Schwarzschild metric. This 
raises the question of whether such a non-zero quadrupole moment has a characteristic 
influence on the shape of Polish doughnuts, which would allow to distinguish such 
impostors from black holes. This is the question we want to investigate in this paper. 

To that end, we construct Polish doughnuts in two spacetimes which
describe the exterior of a source with a non-zero quadrupole moment. 
Both spacetimes are static, axisymmetric and asymptotically flat solutions 
to Einstein's vacuum field equation, i.e., they describe the vacuum region 
around a non-rotating and isolated central object. The first one is the 
so-called $q$-metric, which, in the form considered here, was found by 
Quevedo \cite{Quevedo2011}; the second one is the 
Erez-Rosen metric \cite{bib:er-orig}. These metrics do not feature
a horizon, i.e., they do not describe the spacetime around a black 
hole, unless in the case of a vanishing quadrupole moment where they
reduce to the Schwarzschild metric. If extended to sufficiently low 
radii, they rather display a naked singularity. As many physicists 
consider such naked singularities as unphysical, one may replace the 
metric in an inner region of the spacetime by an interior (e.g. 
perfect-fluid) solution to Einstein's field equation, and then match it 
to the exterior vacuum spacetime at a surface arbitrarily close to the 
naked singularity. In this way, both the $q$-metric and the Erez-Rosen 
metric can be interpreted as viable models for the spacetime around an 
object with a quadrupole moment that may be very compact but not compact 
enough for having collapsed to a black hole. With this interpretation, 
these two metrics make very good black-hole impostors.

The paper is organised as follows. In section \ref{sec:axistat}, we review
relevant background material for constructing Polish doughnuts in an
arbitrary axisymmetric and static spacetime. In section \ref{sec:qmetric},
we specify this to the $q$-metric, and in section \ref{sec:ErezRosen}, we
specify it to the Erez-Rosen metric. It is our main goal to find out
whether or not the accretion tori in these two spacetimes have common features
that distinguish them from accretion tori around a non-rotating and
isolated black hole, i.e., from accretion tori in the Schwarzschild
metric. 

We use geometrised units ($c=G=1$) and our convention for the metric signature 
is $(-, +, +, +)$. Greek indices are running from 0 to 3 and Einstein's summation 
convention is applied.  As usual, we raise and lower indices with the spacetime
metric.

\section{Circular motion in static axisymmetric spacetimes}\label{sec:axistat}

We consider a static axisymmetric metric,
\begin{eqnarray}
    g_{\mu \nu} dx^{\mu} dx^{\nu}  = g_{tt} dt^2  + g_{rr} dr^2 + 
    g_{\vartheta \vartheta} d\vartheta ^2  + g_{\varphi \varphi} d\varphi^2
    \, ,
    \label{eq:ch1:general_metric_vzc}
\end{eqnarray}
where the metric coefficients $g_{\mu \nu}$ are independent of $t$ and $\varphi$. 
We denote the corresponding Killing vector fields by 
\begin{eqnarray}
    \eta^\mu = \delta ^{\mu} _t, \ \ \ \ \ \  \xi^\mu = \delta ^{\mu} _{\varphi} \, .
\end{eqnarray}
We restrict to the part of the spacetime where $g_{\varphi \varphi} >0$ and 
$g_{tt} < 0$. 

In this preparatory section, we want to summarise some properties of matter particles that
are in circular (in general non-geodesic) uniform motion about the symmetry axis of the spacetime.
Then the four-velocity $u^{\mu} \partial _{\mu}$ has to satisfy $u^r = u^\vartheta = 0$ and 
can be written as  
\begin{eqnarray}
    u^\mu = u^t (\eta^\mu + \Omega \xi^\mu),
    \nonumber
\end{eqnarray}
where 
\begin{eqnarray}
\Omega = \dfrac{u^{\varphi}}{u^t}=\frac{\textrm{d} \varphi}{\textrm{d} t}\end{eqnarray}
is the coordinate angular velocity. By uniformity, $\Omega$ 
depends only on $r$ and $\vartheta$. From the normalisation condition 
$g_{\mu \nu} u^\mu u^\nu = -1$, we find
\begin{eqnarray}
    (u^t)^{-2} = - g_{tt} - \Omega^2 g_{\varphi \varphi}
    \, . 
    \label{eq:ch1:four-velocity-omega}
\end{eqnarray}
As $(u^t)^{-2}$ must be positive, this equation restricts $\Omega$ as a function of $r$ 
and $\vartheta$.

We may also characterise the motion by the \emph{specific angular momentum} 
\begin{eqnarray}
l = - \dfrac{u_{\varphi}}{u_t} = - \dfrac{g_{\varphi \varphi} u^{\varphi}}{g_{tt} u^t} \, .
\label{eq:l}
\end{eqnarray}
It is related to the angular velocity $\Omega$ by the equation
\begin{eqnarray}
    g_{\varphi \varphi} \, \Omega = - g_{tt} \, l \, ,
\end{eqnarray}
so we can prescribe the motion either by specifying $\Omega$ as a function of $r$ and 
$\vartheta$, or equivalently by specifying $l$ as a function of $r$ and $\vartheta$. 
If rewritten in terms of
$l$, Eq. \eqref{eq:ch1:four-velocity-omega} reads
\begin{eqnarray}
    u_t^{-2} = - g^{tt} - l^2 \, g^{\varphi \varphi}  \, .
    \label{eq:ut-2} 
\end{eqnarray}
As $u_t^{-2}$ is positive, this equation restricts the values of $l$ that 
are allowed at a certain point ($r$, $\vartheta$).

For the visualisation of circular motion in an axisymmetric and static
spacetime, one often uses the so-called \emph{von Zeipel cylinders}, which are defined
as the surfaces of constant \emph{von Zeipel radius} $\mathcal{R}$,
\begin{eqnarray}
    \mathcal{R}^2 = \dfrac{l}{\Omega}=
    - \dfrac{g_{\varphi \varphi}}{g_{tt}} \label{eq:vzc_general} \, .
\end{eqnarray}
In flat spacetime, where $g_{tt}=-1$ and $g_{\varphi \varphi} = r^2 \mathrm{sin} ^2 \vartheta$, 
they reduce to the ordinary Euclidean cylinders, $\mathcal{R}{}^2= r^2 \mathrm{sin} ^2 \vartheta$.
The relativistic notion of von Zeipel cylinders was introduced by 
Abramowicz \cite{bib:vzc}, who showed that these cylinders are always perpendicular 
to the direction of the centrifugal acceleration. Remarkably, they are the same for \emph{all}
circular motions, independent of the angular velocity. The fact that
the left-hand side of Eq. \eqref{eq:ut-2} must be positive can now be written as
\begin{eqnarray}
    l^2 < \mathcal{R}^2 \, ,
    \label{eq:lR}
\end{eqnarray}
i.e., the region where circular motion is possible with a certain
specific angular momentum $l$ is bounded by the von Zeipel cylinder 
$\mathcal{R}^2 = l^2$.  

\subsection{Effective potential and geometrically thick tori}
\label{sec:geo-thick-tori}

We now recall some basic facts about perfect fluids in circular motion on 
an axisymmetric static spacetime. For details and derivations, we refer to 
Rezzolla and Zanotti \cite{bib:rel-hydro}. Using the angular velocity $\Omega$
and the specific angular momentum $l$ introduced above, the 
Euler equation for a perfect fluid in the case of circular motion reduces to 
\begin{eqnarray}
    \partial_\mu \ln |u_t| - 
    \biggl( \frac{\Omega}{1 - \Omega l } \biggr) \partial_\mu l 
    = 
    - \frac{1}{\rho h} \partial_\mu p
    \label{eq:ch1:euler-equation}
\end{eqnarray}
where $\rho$ is the energy density, $p$ is the pressure and $h$ is the specific enthalpy.

Here we assume that the fluid shares the symmetry of the spacetime, which implies 
that $u_t$, $\Omega$, $l$, $\rho$, $p$ and thus $h$ are functions of $r$ and $\vartheta$
only. For a barotropic fluid, $\rho$, and thus $h$, are functions of $p$, so the the right-hand 
side of \eqref{eq:ch1:euler-equation} becomes a differential. By differentiating both sides, 
it becomes apparent that the differentials $\mathrm{d} l$ and $\mathrm{d} \Omega$ are 
linearly dependent. This implies that either $\mathrm{d} l =0$ or $\Omega = \Omega(l)$. 
This result is known as the (relativistic) \emph{von Zeipel theorem}. 

If $\mathrm{d} l$ has no zeros, the Euler equation \eqref{eq:ch1:euler-equation} can be 
integrated to \cite{bib:vzc}
\begin{eqnarray}
    \mathcal{W} - \mathcal{W}_{\mathrm{in}} := - \int_0^p \frac{\textrm{d} p'}{\rho h} = 
    \ln |u_t| - \ln |(u_t)_{\mathrm{in}}| - \int^l_{l_{\mathrm{in}}} \frac{\Omega}{1 - \Omega l'} \textrm{d}l'
    \label{eq:ch1:potential_raw}
\end{eqnarray}
where the last integral makes sense because $\Omega$ is a function of the specific 
angular momentum. Here $\mathcal{W}$ is an \emph{effective potential} which is 
defined only up to an arbitrary constant, and the suffix ``$\mathrm{in}$'' stands 
for the value of the corresponding quantity at some chosen point; for the construction
of perfect-fluid tori one usually chooses this point at the inner edge of the torus.

If, on the other hand, $\mathrm{d} l$ is identically zero, i.e, if we 
consider a barotropic fluid with constant specific angular momentum, then 
Eq. \eqref{eq:ch1:potential_raw} is also true, but now the last integral
in this equation vanishes. With the help of Eq. \eqref{eq:ut-2}, we then 
find that the effective potential can be expressed as  
\begin{eqnarray}
    \mathcal{W}(l, r, \vartheta ) = 
    - \dfrac{1}{2} \, 
    \mathrm{ln} \Big( -g^{tt} ( r , \vartheta ) 
    - l^2 \, g^{\varphi \varphi} ( r , \vartheta ) \Big)
    \label{eq:eff_pot_general} 
\end{eqnarray}
where we have chosen the arbitrary additive constant in $\mathcal{W}$ 
such that $\mathcal{W}_{\mathrm{in}} = \ln |(u_t)_{\mathrm{in}}|$. 
Equation \eqref{eq:eff_pot_general} determines circular motion of a perfect fluid 
with constant specific angular momentum $l$. Stationarily rotating perfect-fluid
tori with this property are known as ``Polish doughnuts''. They have been discussed 
first by Jaroszy{\'n}ski, Abramowicz and Paczy{\'n}ski in Schwarzschild and Kerr 
spacetimes \cite{bib:donut-ss-kerr}; for a detailed review of the topic, we refer 
again to Rezzolla and Zanotti \cite{bib:rel-hydro}.

The relation between the potential $\mathcal{W}$, the density $\rho$ and the pressure $p$
can be calculated explicitly if we assume a polytropic equation of state 
\begin{eqnarray}
p (r , \vartheta )= K \, \rho (r , \vartheta ) ^ \Gamma \, , \quad 
h(r, \vartheta )= 1+\dfrac{K \, \Gamma \, \rho (r, \vartheta) ^{\Gamma-1}}{\Gamma-1} \label{eq:polytropic}
\end{eqnarray}
where $\Gamma >1$ and $K>0$ are constants, known as the polytropic index and the polytropic constant, 
respectively. Then we can evaluate the integral after the first equality sign in Eq. \eqref{eq:ch1:potential_raw}. 
This gives $\rho$, and thereby $p$, explicitly in terms of the effective potential $\mathcal{W}$,
\begin{eqnarray}
    \rho (r, \vartheta ) = \biggl[ \biggl( \frac{\Gamma - 1}{K \Gamma}\biggr) 
    \biggl( e ^ {\mathcal{W}_{\textrm{in}} - \mathcal{W}(l, r, \vartheta)} - 1 \biggr) \biggr]^{\frac{1}{\Gamma - 1}}
    \, . \label{eq:density-general}
\end{eqnarray}
Note that there are two parameters, $l$ and $\mathcal{W}{}_{\mathrm{in}}$, 
which we can choose freely. As the density $\rho$ must be non-negative, the 
fluid body is restricted to the region where $\mathcal{W}(l,r, \vartheta) \leq 
\mathcal{W}_{\textrm{in}}$. The equation $\mathcal{W}(l,r, \vartheta) = 
\mathcal{W}_{\textrm{in}}$ determines the boundary of the fluid body. By 
fixing $l$, we specify the potential surfaces $\mathcal{W}= \mathrm{constant}$, 
and by fixing the value of $\mathcal{W}{}_{\mathrm{in}}$, we single out a particular 
equipotential surface as the boundary of the fluid body. Note that the 
parameters that enter into the equation of state have no influence
on the shape of the fluid body. They do, however, influence the distribution
of density and pressure in the fluid.


We are interested in the case that $l$ and $\mathcal{W}_{\mathrm{in}}$ 
have been chosen such that the equation $\mathcal{W}(l, r , \vartheta ) =
\mathcal{W}{}_{\mathrm{in}}$ describes a simple closed curve in the $r$-$\vartheta$-plane, 
which means that the fluid occupies the interior of a torus. This is the situation 
the name ``Polish doughnut'' refers to. In this situation, the potential 
$\mathcal{W}(l, r , \vartheta )$, the density $\rho ( r , \vartheta )$ and the
pressure $p(r , \vartheta )$ must have a critical point at 
some $(r_{\mathrm{cen}},\vartheta _{\mathrm{cen}})$ inside the fluid,
called the ``centre'' of the doughnut. If we restrict to the case
that the density, and consequently the pressure, varies monotonically between the
centre and the boundary, this critical point must be a local maximum for $\rho$
and $p$, i.e., a local minimum for $\mathcal{W}$. In the spacetimes considered 
below, we will have 
\begin{eqnarray}
\dfrac{\partial \mathcal{W} ( l , r , \vartheta )}{\partial \vartheta } 
\Bigg| _{\vartheta = \pi /2} = 0 \, , \quad 
\dfrac{\partial ^2 \mathcal{W} ( l , r , \vartheta )}{\partial \vartheta ^2} 
\Bigg| _{\vartheta = \pi /2} > 0 \, .
\label{eq:Wtheta}
\end{eqnarray}
Then a critical point of the potential in the equatorial plane is characterised by
the equation
\begin{eqnarray}
    \Bigg[\dfrac{\partial g^{tt} (r , \vartheta  )}{\partial r} 
    + l^2 \, \dfrac{\partial  g^{\varphi \varphi} (r , \vartheta  )}{\partial r}
    \Bigg]_{\vartheta = \pi/2}  \, = \, 0 \, .
    \label{eq:dW}
\end{eqnarray}
This critical point is a minimum of the potential if 
\begin{eqnarray}
    \pm \Bigg[\dfrac{\partial ^2 g^{tt} (r ,\vartheta )}{\partial r^2} 
    + l^2 \, \dfrac{\partial ^2 g^{\varphi \varphi} (r ,\vartheta )}{\partial r^2} \Bigg]
    _{\vartheta = \pi /2} \, > \, 0
    \label{eq:ddW}
\end{eqnarray}
holds with the upper sign, and it is a saddle if this inequality holds with the lower
sign. So, for the construction of Polish doughnuts, we will have to choose $l$ such that 
there is a point $(r_{\mathrm{cen}}, \vartheta _{\mathrm{cen}} = \pi /2)$ where 
\eqref{eq:dW} and \eqref{eq:ddW} with the upper sign hold.

The fact that the differential of $p$ vanishes at the centre implies that
a fluid element at $(r_{\mathrm{cen}}, \vartheta _{\mathrm{cen}})$ moves 
along a circular timelike geodesic. This follows immediately from the 
fact that, quite generally, the Euler equation reduces to the geodesic
equation at points where $dp$ vanishes. For this reason, circular geodesics 
play an important role in the construction of Polish doughnuts, so we will 
now briefly discuss some of their properties.

\subsection{Circular geodesics}

As before we consider a static and axisymmetric 
spacetime, see Eq. (\ref{eq:ch1:general_metric_vzc}), but now we
assume in addition that the spacetime is symmetric with respect 
to the equatorial plane $\vartheta = \pi /2$. Then a geodesic starting 
tangentially to this plane will remain in this plane. In this section 
we want to calculate the circular timelike and lightlike geodesics 
in the equatorial plane; their properties will become important later. 
To that end we recall that geodesics in the equatorial plane derive 
from the Lagrangian
\begin{eqnarray}
    \mathcal{L} = 
    \dfrac{1}{2} 
    \Big( 
    g_{tt} \dot{t}{}^2 + g_{rr} \dot{r} {}^2 + g_{\varphi \varphi} \dot{\varphi}{}^2
    \Big)
\end{eqnarray}
where the metric coefficients depend on $r$ only, as $\vartheta=\pi/2$. In the following we
write $g_{\mu \nu} (r)$ instead of 
$g_{\mu \nu } ( r , \vartheta = \pi/2 )$ for the sake of brevity.

There are three constants of motion
\begin{eqnarray}
    E = - g_{tt}(r) \dot{t} \, , \quad 
    L = g_{\varphi \varphi}(r) \dot{\varphi} \, , \quad
    \varepsilon = 
    -g_{tt}(r) \dot{t}{}^2 - g_{rr}(r) \dot{r} {}^2 - g_{\varphi \varphi}(r) \dot{\varphi}{}^2
    \label{eq:ELepsilon}
\end{eqnarray}
where $\varepsilon =0$ for lightlike and $\varepsilon = 1$ for timelike geodesics.
The first two equations imply
\begin{eqnarray}
l = \dfrac{L}{E}
\end{eqnarray}
where $l$ is the specific angular momentum defined in \eqref{eq:l}.
Combining the three equations of \eqref{eq:ELepsilon} results in
\begin{eqnarray}
-g_{tt}(r) g_{rr}(r) \dot{r}{}^2 +\mathcal{V} (\varepsilon, L,r) = E^2
\label{eq:con}
\end{eqnarray} 
with the effective potential
\begin{eqnarray}
\mathcal{V} (\varepsilon, L,r) = 
- g_{tt} (r) \Bigg( \dfrac{L^2}{g_{\varphi \varphi} (r) } + \varepsilon \Bigg)   \, .  
\label{eq:Veff}
\end{eqnarray}
Circular geodesics have to satisfy $\dot{r} =0$ and $\ddot{r}=0$, hence
\begin{eqnarray}
\mathcal{V} (\varepsilon, L,r) = E^2 \, , \quad 
\dfrac{\partial \mathcal{V} (\varepsilon, L,r)}{\partial r} = 0      
\, .
\label{eq:circ}
\end{eqnarray}
For lightlike geodesics, $\varepsilon =0$, these two conditions are
equivalent to 
\begin{eqnarray}
\dfrac{L^2}{E^2} = \dfrac{g_{\varphi \varphi} (r)}{-g_{tt}(r)} \, , \quad
g_{tt}(r) g_{\varphi \varphi}' (r) = g_{\varphi \varphi} (r)  g_{tt}'(r)
\, ,
\label{eq:lightcirc}
\end{eqnarray}
where a prime denotes derivative with respect to $r$.
Comparison with \eqref{eq:vzc_general} demonstrates that circular lightlike 
geodesics (``photon circles'') are located precisely at those points in 
the $r$-$\vartheta$-plane where the differential of the von Zeipel radius 
$\mathcal{R}$ vanishes.

For timelike geodesics, $\varepsilon =1$, solving the two equations 
\eqref{eq:circ} for $L$ and $E$ gives us the so-called 
\emph{Keplerian}\footnote{When referring to general-relativistic orbits, 
the attribute ``Keplerian'' means ``relating to timelike circular geodesics''.
} 
values of these constants of motion,
\begin{eqnarray}
L_K (r) ^2 = 
\dfrac{
g_{\varphi \varphi}  (r)^2 g_{tt}'(r)
}{
g_{tt}(r) g_{\varphi \varphi}' (r) - g_{\varphi \varphi} (r)  g_{tt}'(r)
}
\end{eqnarray}
and
\begin{eqnarray}
E_K (r) ^2 = 
\dfrac{
- g_{tt}  (r)^2 g_{\varphi \varphi}'(r)
}{
g_{tt}(r) g_{\varphi \varphi}' (r) - g_{\varphi \varphi} (r)  g_{tt}'(r)
}
\, .
\end{eqnarray}
Their quotient is the \emph{Keplerian specific angular momentum}  
\begin{eqnarray}
l_K (r) ^2 = \dfrac{L_K(r) ^2}{E_K(r)^2} =
\dfrac{
-g_{\varphi \varphi} (r)^2 g_{tt}'(r)
}{
g_{tt}  (r)^2 g_{\varphi \varphi}'(r)
}
=
\dfrac{
-(g^{tt}){}'(r)
}{
(g^{\varphi \varphi}){}'(r)
}
\, .
    \label{lK}
\end{eqnarray}
Analogously, evaluating the angular velocity $\Omega = \dot{\varphi}/\dot{t}$
along circular timelike geodesics gives the \emph{Keplerian angular velocity}
\begin{eqnarray}
    \Omega _K (r) ^2 = \biggl( \dfrac{g_{tt} (r) L_K(r)}{g_{\varphi \varphi} (r) E_K(r)} \biggr)^2
    = \dfrac{-g_{tt}' (r)}{g_{\varphi \varphi} ' (r)}  \, .
    \label{eq:ksam_general}
\end{eqnarray}
A timelike circular geodesic is stable with respect to radial perturbations      
if it is a local minimum of $\mathcal{V}$, and is unstable if it is
a local maximum. For distinguishing these cases, we assume that \eqref{eq:circ}
holds with $\varepsilon = 1$ and calculate
\begin{eqnarray}
\dfrac{\partial ^2 \mathcal{V} (1, L , r)}{\partial r^2} \Bigg| _{L = L_K (r)} 
=
- E_K(r) ^2 g_{tt} (r) 
\Big( (g^{tt}){}'' (r) + l_K(r) ^2 \, (g^{\varphi \varphi}){}'' (r) \Big)
\, .
\label{eq:ddV}
\end{eqnarray}
This demonstrates that a circular timelike geodesic at radius $r$ is stable if the
right-hand side of Eq. \eqref{eq:ddV} is positive, and unstable if it is negative. This
observation has an important consequence in view of the construction of Polish 
doughnuts. If we assume that Eq. \eqref{eq:Wtheta} is satisfied, comparison of Eqs. 
\eqref{lK} and \eqref{eq:ddV} with \eqref{eq:dW} and \eqref{eq:ddW} show the
following: Every circular timelike geodesic in the equatorial plane gives a critical 
point of the potential $\mathcal{W}$. If the geodesic is stable with respect to radial
perturbations, this critical point is a minimum of $\mathcal{W}$; if the geodesic is 
unstable, it is a saddle. This implies in particular that Polish doughnuts can 
exist only around a stable timelike circular geodesic that serves as the centre.

In this paper, we are interested in spacetimes that are 
asymptotically flat with a positive ADM mass $M_0$. Then we have 
$-g_{tt} (r) = 1- 2M_0/r + \mathcal{O}\big( (M_0/r)^2 \big)$ 
with $M_0 >0$ and $g_{\varphi \varphi} (r)/r^2 \to 1 $  
for $r \to \infty$. In this situation, the effective 
potential $\mathcal{V} (1, L , r )$ approaches
the value 1 from below for $r \to \infty$. This implies
that, for sufficiently big $r$, circular timelike geodesics
are stable. At a certain radius $r = r_{\mathrm{ms}}$ they 
may become unstable. At this radius, the 
right-hand side of \eqref{eq:ddV} must be zero, i.e.
\begin{eqnarray}
l_K (r_{\mathrm{ms}}) ^2 = 
\dfrac{-(g^{tt})''(r_{\mathrm{ms}})}{(g^{\varphi \varphi})'' (r_{\mathrm{ms}})}
\, .
\label{eq:ms1}
\end{eqnarray}
By \eqref{lK}, this condition is equivalent to 
\begin{eqnarray}
l_K'(r_{\mathrm{ms}}) = 0 \, .
\label{eq:ms2}
\end{eqnarray}
Eq. \eqref{eq:ms2} is the defining equation of a \emph{marginally stable} 
circular orbit. Note that there may be more than one marginally stable orbit,
i.e., when moving inwards we may encounter again stable orbits after having
crossed a region with unstable ones.

If we have passed the first marginally stable orbit, coming from $r=\infty$, 
we are in a region where $\mathcal{V} (1, L_K(r) , r)< 1$, implying
that a perturbation of an unstable circular orbit in this region 
gives a bound orbit, i.e., an orbit that does not escape to infinity.
The last circular timelike geodesic where this
holds true is called the \emph{marginally bound} orbit. Its radius 
$r_{\mathrm{mb}}$ must satisfy the equation 
$\mathcal{V} (1, L_K (r_{\mathrm{mb}}) , r_{\mathrm{mb}} )=1$,
which can also be rewritten as
\begin{eqnarray}
E_K(r_{\mathrm{mb}}) ^2 = 1  \, .
\label{eq:mb} 
\end{eqnarray}

\section{Tori in the $q$-metric}\label{sec:qmetric}

The most general axisymmetric and static solution of the vacuum field equations is 
represented by the Weyl class of solutions. The simplest of these solutions is the 
Schwarzschild metric, for which all multipole moments but the monopole moment (ADM mass) vanish. 
By the Jebsen-Birkhoff theorem, there is no other Weyl solution for which all the 
higher-order multipole moments vanish. Among the many different Weyl solutions with a
non-vanishing quadrupole moment, the so-called $q$-metric is considered the simplest one. 
It can be written as  
\begin{eqnarray}
    g_{\mu \nu} dx^{\mu} dx^{\nu} 
    = - \biggl( 1 - \frac{2M}{r}\biggr)^{1+q} dt^2 
    \nonumber
\\
\fl  \qquad  + 
    \biggl( 1 - \frac{2M}{r}\biggr)^{-q} 
    \biggl[ \biggl( 1 + \frac{M^2 \sin^2 \vartheta}{r^2 - 2Mr} \biggr)^{-q(2+q)} 
    \biggl( \frac{dr^2}{1 - \frac{2M}{r}} + r^2 d\vartheta^2 \biggr) + 
    r^2 \sin^2 \vartheta d\varphi^2 \biggr] \label{eq:metric_qm}
\end{eqnarray}
where $M$ is a parameter with the dimension of a length and $q$ is a dimensionless parameter.
We refer to $M$ as to the mass parameter and to $q$ as to the quadrupole parameter. In the
representation given here, with the quadrupole parameter $q$, the metric \eqref{eq:metric_qm} 
was found by Quevedo \cite{Quevedo2011}, cf. Quevedo et al. \cite{bib:quevedo2013quadrupolar}. 
In other representations, this solution to the Einstein vacuum field equation was known before; 
it is a special case of a class of metrics discussed by Bach \cite{Bach1922} and it was also 
studied e.g. by Zipoy \cite{Zipoy1966} and Vorhees \cite{Voorhees1970}. We also mention that 
Toktarbay and Quevedo \cite{ToktarbayQuevedo2014}  found a rotating (non-static) 
generalisation of the $q$-metric which, however, will not be considered in the following. 

For $q=0$ the $q$-metric reduces to the spherically symmetric Schwarzschild 
metric with ADM mass $M$. For $q=-2$ it also reduces to the Schwarzschild metric,
as can be seen by transforming the radius coordinate according to $r \mapsto r-2M$;
this time, however, the ADM mass is equal to $-M$ if we define it, as usual, at
$r = + \infty$. For any values of $q$ and $M$ the 
$q$-metric is symmetric with respect to the equatorial plane 
$\vartheta = \pi/2$ and asymptotically flat. For $M = 0$ and any value of $q$, it is flat.
The same is true for $q=-1$ and any value of $M$, which is less obvious.
In the case $M > 0$ and $q \neq -1$, the metric \eqref{eq:metric_qm} is singular
at the positive radius value $r = 2M$. While in the Schwarzschild case $q=0$ this 
is a mere coordinate singularity, indicating a horizon, there is a naked curvature 
singularity at $r=2M$ for all other values of $q$. As naked singularities are widely 
believed to be unphysical, it is reasonable to consider the $q$-metric only outside 
of a closed surface $\mathcal{S}$ that surrounds the naked singularity, and to think 
of an interior matter solution being matched to this metric at $\mathcal{S}$, see 
Stewart et al. \cite{bib:StewartEtAl1982} and Quevedo et al. \cite{bib:quevedo2013quadrupolar}. 
With this interpretation, the $q$-metric describes the gravitational field outside of a 
deformed celestial body that might be quite compact, but not compact enough to form a black 
hole.

The Geroch-Hansen mass monopole and quadrupole moments of the $q$-metric can 
be expressed as \cite{Quevedo2011}
\begin{eqnarray}
    M_0 = (1 + q)M, \ \ \ M_2 = - \frac{M^3}{3} q (1 + q) (2 + q) \, .
    \label{eq:M2_qm}
\end{eqnarray}
All even multipole moments are different from zero and determined
by $M$ and $q$. 

As we want the ADM mass $M_0$ to be positive, we have to
choose either $M>0$ and $q>-1$ or $M<0$ and $q<-1$. Actually, 
it suffices to consider the first case. The reason is that the
$q$-metric with parameters $M$ and $q$ is isometric to the 
$q$-metric with parameters $M'=-M$ and $q'=-q-2$, as can be 
seen by transforming the radius coordinate according to 
$r \mapsto r' = r - 2M$. Therefore, the $q$-metrics with
$M>0$ and $q>-1$ are the same as the ones with $M'<0$
and $q'<-1$. 

From now on we assume that $M>0$ and $q>-1$.  
For negative values of $q>-1$, the mass quadrupole moment $M_2$ 
is positive; the latter property means that, far away from 
the centre, the two-dimensional surfaces $(g_{tt} = \mathrm{const.}, 
t= \mathrm{const.})$ are prolate spheroids. They are oblate spheroids
for positive values of $q$, which corresponds to negative values of
$M_2$, cf. Quevedo et al. \cite{bib:quevedo2013quadrupolar, bib:multipole-moments}.

For better comparison with other (quadrupolar) spacetimes, it is convenient to 
express the $q$-metric in terms of $M_0$ and $M_2$, rather than in terms of $M$ 
and $q$. The transformation $(M_0,M_2) \mapsto (M,q)$ is given by:
\begin{eqnarray}
    M = M_0 \sqrt{3\frac{ M_2}{M_0^3} + 1}, \ \ \ 
    q = \frac{1}{\sqrt{3\frac{ M_2}{M_0^3} + 1}} - 1  \, .
\end{eqnarray}
Demanding the expression under the root to be positive, resulting in a real 
(and positive) $M$, requires $M_2 / M_0^3 > - 1 / 3$. In other words, in the 
oblate case the values of $M_2$ are restricted once $M_0$ has been chosen. 
We will see below that there is no such restriction for the Erez-Rosen metric. 

For later convenience we also mention that the Kretschmann scalar of the 
$q$-metric can be expressed as \cite{bib:frutosalfaro2017comparison}
\begin{eqnarray}
\fl \qquad   \mathcal{K} = R_{\beta \gamma \delta \eta} R^{\beta \gamma \delta \eta} = 
    \frac{
    16 M^2 (1 + q)^2 \, (r^2 - 2Mr + M^2 \sin^2{\vartheta})^{2(2q + q^2) - 1}
    }{
    r^{4 (2 + 2q + q^2)} \, (1-2M/r)^{2(q^2 + q + 1)} 
    } 
    \; \mathcal{F}(r, \vartheta)
\end{eqnarray}
with
\begin{eqnarray}
\fl \qquad \mathcal{F} ( r , \vartheta ) =&
    3 (r - 2M - qM)^2(r^2 - 2Mr + M^2 \sin^2 \vartheta) \nonumber \\
    &+ M^2 q (2 +q ) \sin^2 \vartheta [M^2 q (2 + q) + 3(r - M) (r - 2M - qM)] \, .
\end{eqnarray}
In the Schwarzschild case $q=0$, the Kretschmann scalar reduces to $\mathcal{K} = 48 M^2/r^6$,
being regular at $r=2M$. For all other values of $q(>-1)$, the Kretschmann scalar 
diverges at $r=2M$, indicating a curvature singularity.

\subsection{Lightlike and timelike circular geodesics}\label{subsec:geoqm}
Next, lightlike and timelike circular geodesics in the equatorial plane are briefly discussed.
For the $q$-metric, the effective potential \eqref{eq:Veff} specifies to
\begin{eqnarray}
    \mathcal{V} = \biggl( 1- \frac{2M}{r} 
    \biggr)^{q+1} \biggl[ \frac{L^2}{r^2} \, \biggl( 1- \frac{2M}{r} \biggr)^q + \varepsilon \biggr] 
    \label{eq:veff_qm}
\end{eqnarray}
For lightlike geodesics ($\varepsilon =0$), the conditions \eqref{eq:circ} for
circular orbits admit exactly one solution for the radius
\begin{eqnarray}
    r_c = (3+2q) M = \biggl( 2 + \sqrt{1 + 3 \frac{ M_2}{M_0^3} } \biggr)  M_0
    \label{eq:rc}
\end{eqnarray}
and the specific angular momentum 
\begin{eqnarray}
    \dfrac{L^2}{E^2} = \dfrac{r_c^2}{\Big(1-\dfrac{2M}{r_c} \Big)^{1+2q}} 
    \, .
    \label{eq:LE}
\end{eqnarray}
Thus, a photon circle exists in the region outside the singularity
if $q > -1/2$. This photon circle is unstable with respect to radial perturbations.
In the Schwarzschild limit $q \to 0$, \eqref{eq:rc} and \eqref{eq:LE} reduce to 
the well-known values $r_c=3M$ and $L^2/E^2 = 27 M^2$. For positive values of 
$q$ (negative values of $M_2 / M_0^3$), 
the radius $r_c$ is greater than $3 M$ and approaches infinity for $q \to \infty$;
for negative values of $q$ (positive values of $M_2 / M_0^3$), it is smaller than 
$3 M$ and approaches the singularity at $2M$ for $q \to -1/2$. For a plot of $r_c$
as a function of the quadrupole moment, see figure \ref{fig:r_limits_qm}. 

For timelike geodesics ($\varepsilon = 1$), the conditions \eqref{eq:circ} for
circular orbits result in the Keplerian values
\begin{eqnarray}
    L_K (r) ^2 = \biggl( 1- \frac{2M}{r} \biggr)^{-q}  \dfrac{(1+q) M r^2}{r-(3+2q)M}
    \label{eq:L2q}
\end{eqnarray}
and
\begin{eqnarray}
    E_K (r) ^2 = \biggl( 1- \frac{2M}{r} \biggr)^{1+q}  \, 
    \dfrac{ r- (2+q) M}{r-(3+2q)M}
 \,.
\end{eqnarray}
As $L_K(r)^2$ cannot be negative, circular timelike geodesics
exist for $r>r_c$ where $r_c$ is the radius of the unstable circular lightlike 
geodesic given in \eqref{eq:rc}. 

From $L_K(r)$ and $E_K(r)$ we find the Keplerian 
specific angular momentum
\begin{eqnarray}
    l_K (r) ^2 = \biggl( 1- \frac{2M}{r} \biggr) ^{-(1+2q)} \dfrac{ (1+q) M \, r^2}{r - (2 + q)M} 
    \label{eq:ksam_qm}
\end{eqnarray}
and, with (\ref{eq:ksam_general}), the Keplerian angular velocity
\begin{eqnarray}
    \Omega_K (r) ^2= \biggl( 1- \frac{2M}{r} \biggr) ^{1+2q}  \frac{(1+q) M}{r^2 (r - ( 2 + q )M)} \, .
    \label{eq:ang_freq_qm}
\end{eqnarray}

For determining the marginally bound orbits, we have to solve the equation 
$E_K(r)^2= \mathcal{V} (1, L_K(r),r) = 1$ for $r$. This cannot be done analytically, 
thus $r_{\mathrm{mb}}$ is only computed numerically. The result is depicted  
in figure \ref{fig:r_limits_qm}. 

The marginally stable circular orbits in the $q$-metric have been calculated
already by Boshkayev et.al. \cite{bib:orbits-$q$-metric}. In general, there are
two of them, an inner one at $r=r_{\mathrm{ms}}^-$ and an outer one at 
$r=r_{\mathrm{ms}}^+$, where
\begin{eqnarray}
    r_{\mathrm{ms}} ^\pm  = M \biggl(4 + 3q \pm \sqrt{5q ^2 + 10 q + 4} \biggr) 
    \, .
    \label{eq:rms_qm}
\end{eqnarray}
This result can easily be verified by solving the  
equation \eqref{eq:ms2}. When evaluating Eq. \eqref{eq:rms_qm}, 
one has to keep in mind that $q > -1$ and $r>r_c$. These restrictions 
divide the family of $q$-metrics into three classes:

\begin{figure}[t!]
    \centering
    \includegraphics[width=1.1\textwidth]{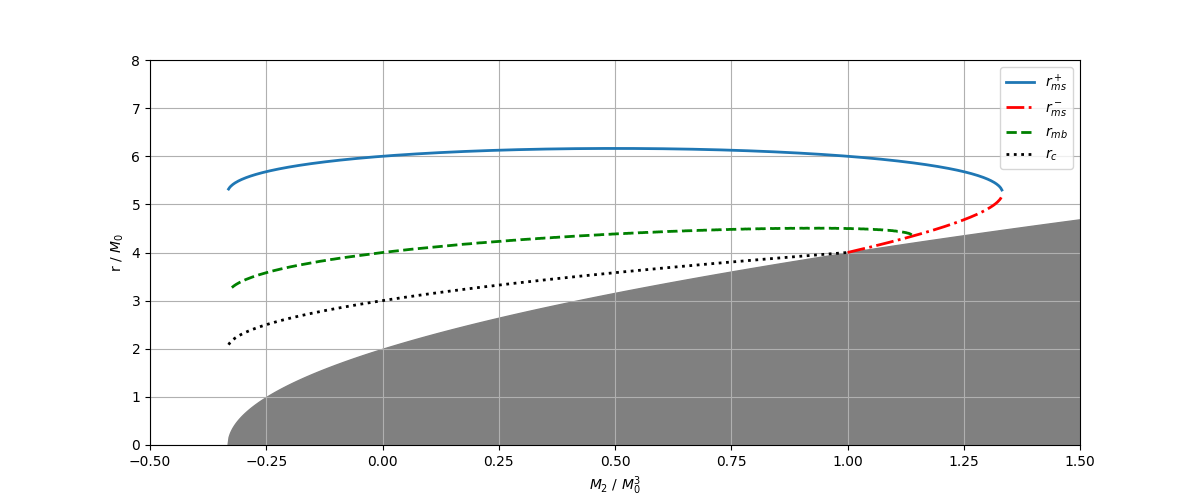}
    \caption{Circular equatorial geodesics in the $q$-metric, depending on the mass quadrupole moment 
    $M_2 / M_0^3$. Note that $q \to \infty$ corresponds to $M_2/M_0^3 = -1/3$, and $q=-1$ corresponds 
    to $M_2/M_0^3 \to \infty$. Blue solid line: positive solution of the marginally stable circular orbit 
    $r_{\mathrm{ms}}^+$; red dash-dotted line: negative solution of the marginally stable circular orbit 
    $r_{\mathrm{ms}}^-$; green dashed line: marginally bound circular orbit $r_{\mathrm{mb}}$; 
    black dotted line: photon circle $r_c$. The grey shaded area marks the region inside the naked 
    singularity, $r\leq 2M$.}
    \label{fig:r_limits_qm}
\end{figure}

\begin{description}

\item[Class I]: $\infty > q > -1/2$, i.e. $-1/3 < M_2 /M_0^3 < 1$ \\
In this class, there is only one marginally stable orbit at $r_{\mathrm{ms}}^+$. 
Orbits between $r_{\mathrm{ms}}^+$ and infinity are stable, whereas orbits 
between $r_c$ and $r_{\mathrm{ms}}^+$ are unstable. Between $2M$ and $r_c$,
there are no circular timelike geodesics. For $q \to -1/2$, we find 
$r_{\mathrm{ms}}^+ /M \to 3$, which corresponds to $r_{\mathrm{ms}}^+/M_0 \to 6$,
and for $q \to \infty$ we find $r_{\mathrm{ms}}^+/M \to \infty$ which 
corresponds to $r_{\mathrm{ms}}^+/M_0 \to 3 + \sqrt{5}$.

\item[Class II]: $-1/2 > q > -1 + 1/\sqrt{5} \approx -0.553$, i.e., 
$1 < M_2 /M_0^3 < 4/3$ \\
In this class, there are two marginally stable circular orbits at 
$r_{\mathrm{ms}}^\pm$. 
In the interval $r_{\mathrm{ms}}^- < r < r_{\mathrm{ms}}^+$ 
circular orbits are unstable, for radii $r > r_{\mathrm{ms}}^+$ 
and $2M < r < r_{\mathrm{ms}}^-$, they are stable.
In the limiting case $q \to -1 + 1/{\sqrt{5}}$, the two marginally stable
orbits merge at $r_{\mathrm{ms}} ^{\pm} = (1 + 3 / \sqrt{5} ) M = (3 + \sqrt{5}) M_0$.

\item[Class III]: $ -1 + 1/\sqrt{5} > q > -1$, i.e., $4/3 < M_2/M_0^3 < \infty$ \\
In this class, there are no marginally stable circular orbits.
At any radius $r>2M$ there is a stable circular
timelike geodesic. Note that $l_K (r) ^2 \to 0$ and $\Omega _K (r) ^2 \to \infty$
for $r \to 2M$.

\end{description}
The radii of the photon circle, the marginally bound orbit and the marginally
stable orbits, given in units of $M_0$, are plotted in figure \ref{fig:r_limits_qm} 
against the quadrupole moment $M_2 / M_0^3$.
The graphs of $r_c$, $r_{\mathrm{mb}}$ and $r_{\mathrm{ms}}^+$ start 
at $M_2 / M_0^3 = - 1/3$, which corresponds to 
$q \rightarrow \infty$. The photon radius $r_c$ increases monotonically, but 
drops under the position of the singularity at $M_2 / M_0^3 = 1$,
corresponding to $q = -1/2$. 
Above this value of $M_2/M_0^3$, there are two marginally stable circular orbits 
which merge at $M_2 / M_0^3 = 4 / 3$, and then vanish.
Between the photon orbit and the marginally stable orbit, the marginally 
bound circular orbit can be found; it vanishes above $M_2 / M_0^3 \approx 1.13$, 
where the maximum of $\mathcal{V}$ drops below 1.

\begin{figure}[ht!]
    \centering
    \includegraphics[width=1\textwidth]{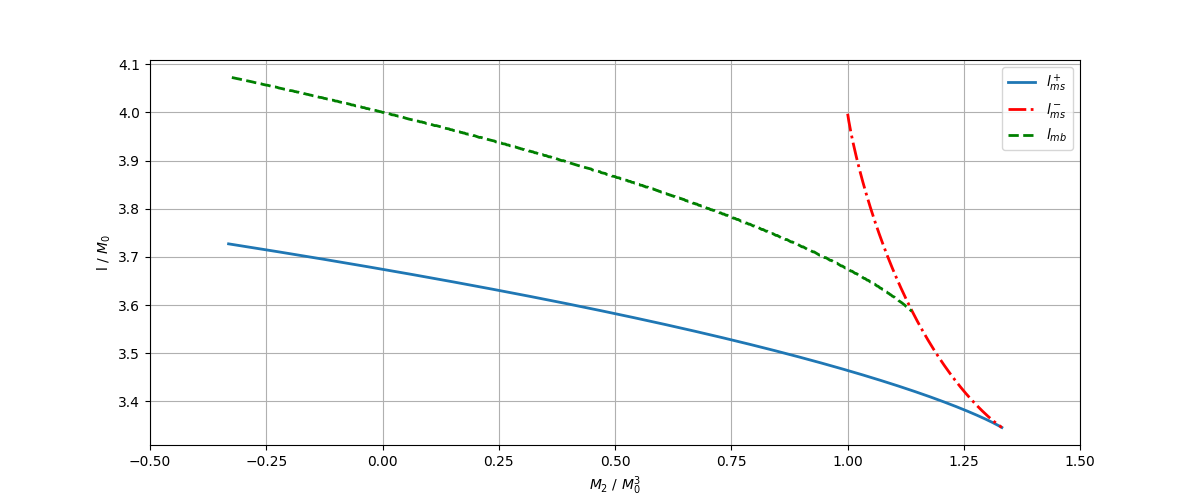}
    \caption{Specific angular momentum of the marginally stable orbits (blue solid and 
    red dash-dotted lines) and of the  marginally bound orbits (green dashed line) in the 
    $q$-metric, depending on the mass monopole moment $M_2 / M_0^3$.}
    \label{fig:limits_qm}
\end{figure}

Inserting the radius values $r_{\mathrm{ms}}^{\pm}$ and $r_{\mathrm{mb}}$ into
\eqref{eq:ksam_qm} gives the specific angular momentum of the marginally 
stable and marginally bound orbits, which we denote by $l_{\mathrm{ms}}^{\pm}$
and $l_{\mathrm{mb}}$, respectively. In figure \ref{fig:limits_qm}, we show 
plots of $l_{\mathrm{ms}}^{\pm}$ and $l_{\mathrm{mb}}$, depending on the 
quadrupole moment. 

\subsection{Von Zeipel cylinders}
\begin{figure}[ht!]
    \centering
    \includegraphics[width=\textwidth]{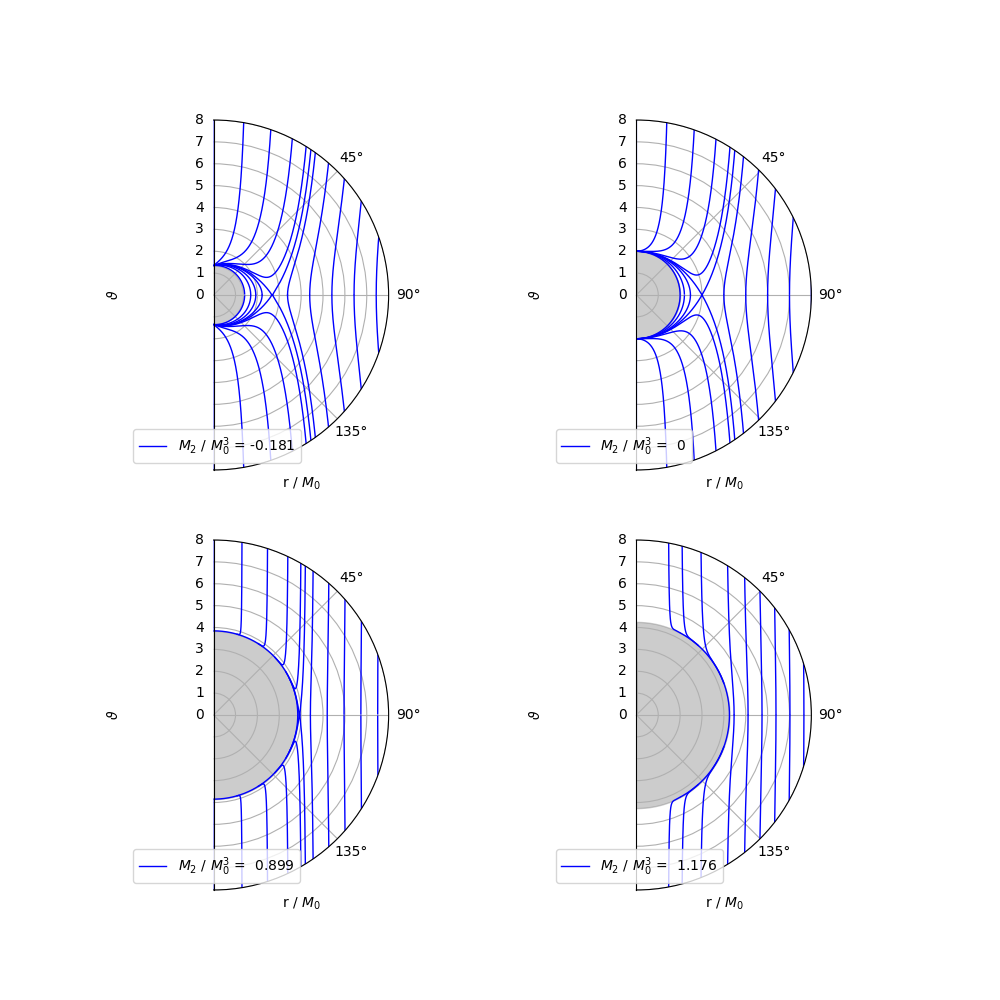}
    \caption{Von Zeipel cylinders for different mass quadrupole moments in the $q$-metric. Upper-left: von Zeipel cylinders for negative values of $M_2 / M_0^3$, corresponding to an oblate geometry; upper-right: von Zeipel cylinders for vanishing quadrupole moment, corresponding to the Schwarzschild limit; lower-left: von Zeipel cylinders for positive values of $M_2 / M_0^3$, corresponding to a prolate geometry; lower-right: von Zeipel cylinders for $M_2 / M_0^3$ in Case II - here, there is no crossing surface anymore, as there is no photon circle. The grey-shaded areas indicate the area below the singularity $r \leq 2M$.}
    \label{fig:vzc_qm}
\end{figure}
In this section, the von Zeipel cylinders are calculated in the $q$-metric.
They give an insight into the behaviour of circular motions in this metric.
Recall that, for any static and axisymmetric spacetime, these cylinders are independent 
of the specific angular momentum.

The von Zeipel cylinders are defined as the surfaces $\mathcal{R} = \mathrm{const.}$ with
$\mathcal{R}$ given by \eqref{eq:vzc_general}. In the $q$-metric, this reduces to
\begin{eqnarray}
\mathcal{R} (r , \vartheta ) ^2 =  \dfrac{\big( 1- \frac{2M}{r} \big)  ^{1+2q}}{r^2 \mathrm{sin} ^2 \vartheta}
\, .
\end{eqnarray}
The surfaces $\mathcal{R} = \mathrm{const.}$ are plotted in 
figure \ref{fig:vzc_qm} for different quadrupole moments $M_2 / M_0^3$. 
Note that, according to \eqref{eq:lightcirc}, the differential 
$\mathrm{d} \mathcal{R}$ has a zero at the photon circle 
$(r=r_c, \vartheta = \pi /2 )$; the corresponding von Zeipel 
cylinder has a self-crossing there.

\subsection{Effective potential}
\begin{figure}[ht!]
    \centering
    \includegraphics[width=1.1\textwidth]{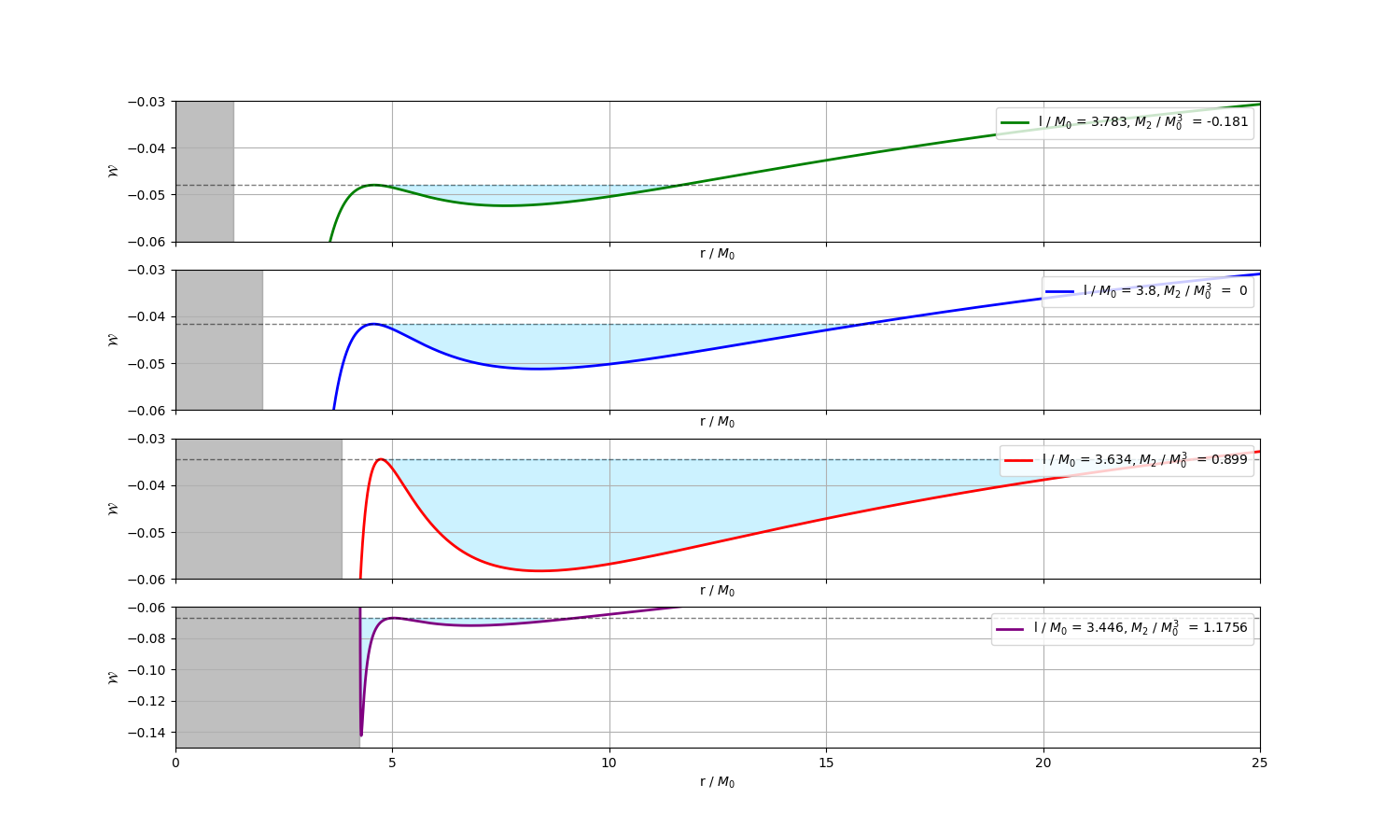}
    \caption{Effective potential
    in the equatorial plane in the $q$-metric, depending on $M_2 / M_0^3$. 
    The grey shaded area indicates the region bounded by the singularity, the 
    lighter blue-shaded area is the region occupied by a torus that fills its
    Roche lobe completely. Upper graph: Class I
    spacetime with a negative value of $M_2 / M_0^3$, corresponding to an oblate geometry; 
    upper-middle graph: Schwarzschild spacetime; 
    lower-middle graph: Class I spacetime with a positive value of $M_2 / M_0^3$, corresponding 
    to a prolate geometry; lower graph: Class II spacetime, note that effective potential approaches $+\infty$
    for $r \to r_{\mathrm{sing}}$.}
    \label{fig:eff_pot_qm}
\end{figure}

\begin{figure}[ht!]
    \centering
    \includegraphics[width=1\textwidth]{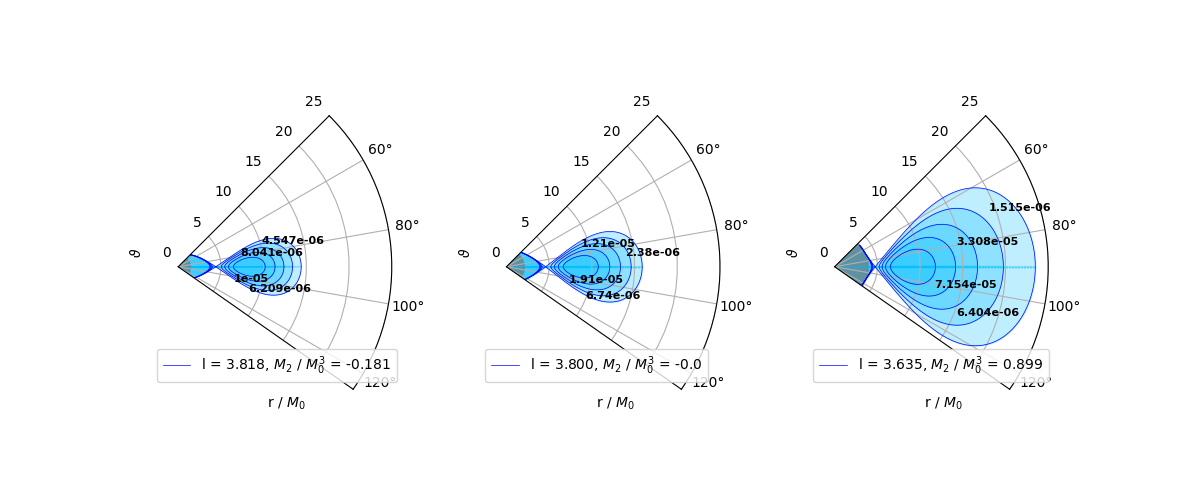}
    \caption{Polar dependency of the effective potential in the $q$-metric. The black numbers represent the
    density in units of $M_0^{-2}$, assuming a polytropic equation of state with $K = 0.2$ and $\Gamma = 5 / 3$.
    We compare the density $\rho$ with the Kretschmann scalar $\mathcal{K}$ at the centre to demonstrate that
    the gravitative influence of the fluid is negligible. 
    Left: torus in a Class I spacetime with a negative value of 
    $M_2 / M_0^3$, corresponding to an oblate geometry, 
    $\rho (r_{\mathrm{cen}})^2 / \mathcal{K}(r_{\mathrm{cen}}) = 3.528 \times 10^{-6} $ ; 
    middle: torus in the Schwarzschild spacetime, 
    $\rho (r_{\mathrm{cen}})^2 / \mathcal{K}(r_{\mathrm{cen}}) =1.18 \times 10^{-6}$; 
    right: torus in a Class I spacetime with a positive value of $M_2 / M_0^3$, corresponding to a prolate
    geometry, $\rho (r_{\mathrm{cen}})^2 / \mathcal{K}(r_{\mathrm{cen}}) = 9.389 \times 10^{-8}$.}
    \label{fig:eff_polar_qm}
\end{figure}

\begin{figure}[ht!]
    \centering
    \includegraphics[width=1.2\textwidth]{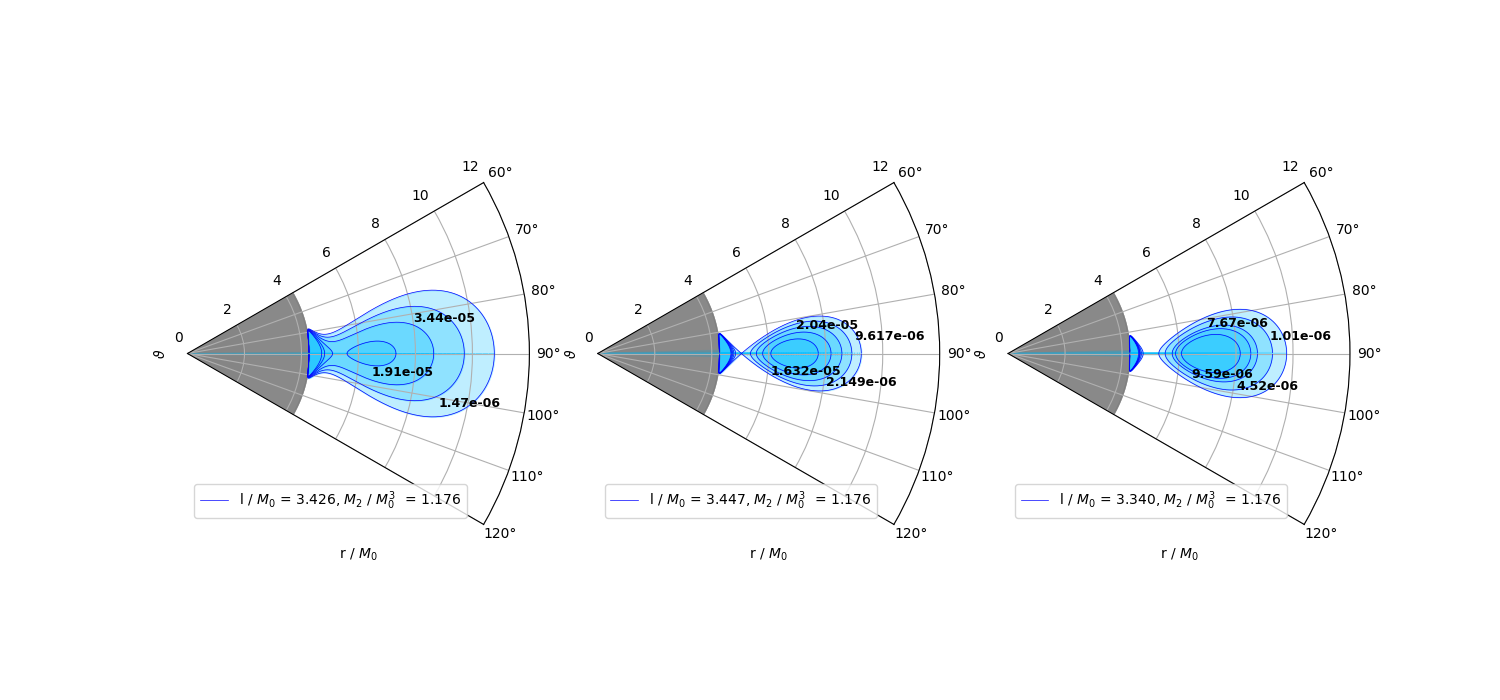}
    \caption{Polar dependency of the effective potential in the $q$-metric, for connected double tori (left panel),
    touching double tori (middle panel) and spatially separated tori (right panel),
    for a spacetime of Class II. The black numbers represent the 
    density where we have assumed a polytropic equation of state with $K = 0.2$ 
    and $\Gamma = 5 / 3$; left: $\rho^2 (r_{\mathrm{cen}}) / \mathcal{K} (r_{\mathrm{cen}}) = 1.78 \times
    10^{-5}$, middle: $\rho^2 (r_{\mathrm{cen}}) / \mathcal{K} (r_{\mathrm{cen}}) = 1.04 \times
    10^{-5}$, right: $\rho^2 (r_{\mathrm{cen}}) / \mathcal{K} (r_{\mathrm{cen}}) = 8.74 \times
    10^{-6}$. Note that the inner Roche lobe is in close proximity to
    the singularity at $r=2M$, but is indeed isolated from it.}
    \label{fig:fish_qm}
\end{figure}

For constructing Polish doughnuts, we calculate the effective 
potential \eqref{eq:eff_pot_general} for the $q$-metric:
\begin{eqnarray}
    \mathcal{W}(l,r, \vartheta) = \frac{1}{2} 
    \ln{\biggl[\frac{
    r^2 \sin^2 \vartheta
    }{
    \big( 1- \frac{2M}{r} \big) ^{- (1+q) } r^2 \sin^2 \vartheta - l^2 \big( 1- \frac{2M}{r} \big) ^q } \biggr]} 
    \, .
    \label{eq:eff_pot_qm}
\end{eqnarray}
In figure \ref{fig:eff_pot_qm} we plot this potential restricted to the equatorial 
plane as a function of the radius coordinate, for various values of $M_2$ and $l$ 
with $M_0$ chosen as the unit 
of length. In figures \ref{fig:eff_polar_qm} and \ref{fig:fish_qm} we show the 
equipotential surfaces $\mathcal{W}= \mathrm{const.}$ in the $r$-$\vartheta$-plane, again 
for various values of $M_2$ and $l$. For the boundary of the torus, we may choose
any closed equipotential surface. We indicate the maximal torus for a given $l$
by a blue-shaded area; this maximal domain is known as the \emph{Roche lobe}.
Recall that the equipotential surfaces and, consequently the shape of the torus, is 
independent of the parameters that enter into the equation of state. We have
also calculated the density $\rho$, which does depend on the equation of state.
In figures \ref{fig:eff_polar_qm} and \ref{fig:fish_qm}, we give the 
density $\rho / M_0^2$, where we have assumed a particular polytropic 
equation of state, as explained in section 2.

We will now discuss the properties of the Polish doughnuts in the $q$-metric 
and the differences to the Schwarzschild case in detail. We recall that,
for the construction of the accretion tori, we need a stable circular timelike geodesic in the 
equatorial plane that can serve as the centre. The region where such 
orbits occur depends on the class of $q$-metrics described above. 
Thus, the three classes are discussed individually. Note that in \emph{any} case the
potential $\mathcal{W}$ goes to 0 from below for $r \to \infty$.

We begin with Class I, which contains the Schwarzschild spacetime as a 
special case. 
Here, we have a photon circle at radius $r_c$, a marginally
bound orbit at radius $r_{\mathrm{mb}}$ and exactly one marginally stable orbit at
$r_{\mathrm{ms}}^+$, where $2M<r_c<r_{\mathrm{mb}}<r_{\mathrm{ms}}^+ < \infty$.
For the construction of the torus, one chooses $l$ such that $l^2>(l_{\mathrm{ms}}^+)^2$;
this guarantees the existence of a centre. Then the potential $\mathcal{W}$ 
features a local minimum at some radius $r_{\mathrm{cen}}$, and a local maximum at 
some radius $r_{\mathrm{cusp}}$, where $r_{\mathrm{cusp}} <  r_{\mathrm{cen}}$; the
potential goes to $- \infty$ for $r \to 2M$, see the first three panels of 
figure~\ref{fig:eff_pot_qm} for examples. Following the definition of the marginally 
bound orbit, it follows that $\mathcal{W}(l,r_{\mathrm{cusp}},\pi/2)<0$
as long as $l^2 < l_{\mathrm{mb}}^2$. In that case, we get a closed equipotential surface
by choosing $\mathcal{W}_{\mathrm{in}}$ such that $\mathcal{W}(l,r_{\mathrm{cen}},\pi/2) < \mathcal{W}_{\mathrm{in}}
\le \mathcal{W}(l,r_{\mathrm{mb}},\pi/2)$. The maximal volume the fluid can occupy without overflowing 
is reached by choosing $\mathcal{W} _{\mathrm{in}} = \mathcal{W}(l, r_{\mathrm{mb}},\pi/2)$; we have already
mentioned that this volume is called the ``Roche lobe''. 
The suffix ``cusp'' in $r_{\mathrm{cusp}}$ refers to the fact that
the Roche lobe features a cusp at $(r=r_{\mathrm{cusp}},\vartheta = \pi /2)$.
In figure \ref{fig:eff_polar_qm}, we show three such tori that fill their Roche lobes 
completely, restricted to the equatorial plane. Tori that fill 
their Roche lobes completely are of particular interest because the fluid is
at the verge of overflowing towards the centre; in this sense, such stationary
rotating tori approximate fluid configurations with an accretion flow onto 
the central object. 
For constructing Polish doughnuts, we may also choose
$l^2 > l_{\mathrm{mb}}^2$. Then $\mathcal{W}_{\mathrm{in}}$ is limited by $\mathcal{W}_{\mathrm{in}}<0$,
as otherwise the torus would extend to infinity. Such a Polish doughnut 
does not feature a cusp at its inner edge, so it is not at the verge of 
accretion. For this reason, tori with $(l_{\mathrm{mb}}^+)^2 < l^2
< l_{\mathrm{mb}}^2$ are usually considered as being more interesting.

We now turn to spacetimes of Class II, which show considerably different 
qualitative features. In this case there are two marginally stable orbits and, potentially, one 
marginally bound orbit in the unstable region between them, $2M< r_{\mathrm{ms}} ^- \ 
(< r_{\mathrm{mb}}) < r_{\mathrm{ms}}^+ < \infty$. Qualitatively different types of Polish 
doughnuts exist for $l^2>(l_{\mathrm{ms}}^+)^2$. Here, the potential 
$\mathcal{W}(l,r, \pi/2)$ has two minima at $r_{\mathrm{cen,1}}$ and $r_{\mathrm{cen,2}}$,
and a maximum at $r_{\mathrm{cusp}}$; the effective potential goes to $+ \infty$ at some radius 
$r_s$ where $2M<r_s < r_{\mathrm{cen,2}} < r_{\mathrm{cusp}} < r_{\mathrm{cen,1}} 
< \infty$, see the last panel of figure  \ref{fig:eff_pot_qm}. 
Regardless of whether the marginally bound orbit exists, 
we may always choose $l$ such that $\mathcal{W}(l,r_{\mathrm{cusp}},\pi/2)<0$. If 
the marginally bound orbit exists, this can be achieved by choosing
$(l_{\mathrm{ms}}^+)^2<l^2<l_{\mathrm{mb}}^2$; if it does not exist, 
any choice $(l_{\mathrm{ms}}^+)^2<l^2$ will do. In that case, the inequality
$\mathcal{W}(l,r,\vartheta )\le \mathcal{W}(l,r_{\mathrm{cusp}}, \pi/2)$ defines \emph{two}
Roche lobes, both isolated from the singularity at $r=2M$ and from infinity,
touching at $(r=r_{\mathrm{cusp}},\vartheta = \pi/2)$. Perfect-fluid configurations
in the form of double tori also exist in the Kerr spacetime, see 
Pugliese et al. \cite{PuglieseMontani2015,PuglieseStuchlik2017}. Double tori
have also been found for \emph{charged} fluids in the presence of a magnetic
field in the Schwarzschild spacetime, see Trova et.al. \cite{trova2020}.
However, they have not been observed before, to the best of our knowledge, 
for uncharged fluids on a static spacetime. In particular, such double tori 
cannot be constructed in the unperturbed Schwarzschild metric, or in any 
$q$-metric of Class I, irregardless of the choice of $l$. 
The last panel in figure \ref{fig:eff_pot_qm} and figure \ref{fig:fish_qm} 
show such double tori. As one can
see in figure \ref{fig:fish_qm}, the cross-section of the Roche lobes has 
the shape of a fish\footnote{\emph{Polish Fishes}, as one might call them}, 
with the outer lobe corresponding to the body of the fish and the inner 
lobe to the fish-tail. The major difference of figure \ref{fig:fish_qm} in
comparison to figure \ref{fig:eff_polar_qm} is in the fact that the ``fish-tail'' is
isolated from the singularity, i.e., we can match at a surface $\mathcal{S}$
an interior solution to the $q$-metric in such a way that the inner torus 
does not touch $\mathcal{S}$. Note that in such double-torus configurations
the inner and the outer torus could rotate in opposite directions if they 
do not fill their Roche lobes completely. As the inner Roche lobe has no 
cusp at its \emph{inner} edge, these double tori are not at the 
verge of accretion. Correspondingly, if the two tori
are bigger than their Roche lobes, the fluid will not spill over towards the
centre or towards infinity; it will rather form a single torus, 
without cusps, but with two centres; such a 
double-centre torus can be constructed, with the same $l$ as the 
double tori, by choosing $\mathcal{W}_{\mathrm{in}}$ such that 
$\mathcal{W}(l,r_{\mathrm{cusp}}, \pi/2) < \mathcal{W}_{\mathrm{in}}<0$, 
see the picture on the left in figure \ref{fig:fish_qm}. This 
corresponds to raising the dashed line in the lowermost panel of 
figure \ref{fig:eff_pot_qm}. 

In spacetimes of Class III, there is a stable circular orbit at any radius
$r>2M$. Each of these orbits can be chosen as the centre of a Polish
doughnut. However, as there are no unstable circular orbits, no cusp
exists, i.e., the Polish doughnuts in such a spacetime can never be 
at the verge of accretion. All tori in this class
are qualitatively similar to the tori with $l^2> l_{\mathrm{mb}}^2$ constructed in 
the Schwarzschild spacetime.

Note that we have assumed that the spacetime geometry is given by the unperturbed
$q$-metric, i.e., we have neglected the influence of the mass of the fluid on the spacetime. T
his assumption is justified as long as the density of the fluid is sufficiently small. For the 
numerical examples shown in figure \ref{fig:eff_polar_qm}, we demonstrate in the
caption that the density squared at the centre of the torus is much smaller
than the Kretschmann scalar at this point, demonstrating that the torus has 
a negligible effect on the spacetime geometry.


\section{Tori in the Erez-Rosen spacetime}\label{sec:ErezRosen}

Next, we consider the \emph{Erez-Rosen} metric, which is a 
static and axisymmetric solution to Einstein's vacuum field 
equation that was found by Erez and Rosen in 1959 \cite{bib:er-orig}. 
It was the first metric found in the class of Weyl solutions that 
was identified as describing the gravitational field around a central 
object with a quadrupole moment. For the correction of several 
misprints in the original paper by Erez and Rosen, we refer to 
Young and Coulter \cite{bib:er-correction}.

The metric can be written in prolate spheroidal coordinates $(t,\varphi,x,y)$ 
as 
\begin{eqnarray}
 \mathrm{d} s ^2 = g_{\mu \nu} \mathrm{d} x^{\mu} \mathrm{d} x^{\nu}  =- f(x,y) \mathrm{d} t^2 \nonumber \\
    + \frac{M^2}{f(x,y)} \biggl[ e^{2 \gamma (x,y)} (x^2 - y^2) 
    \biggl( \frac{\mathrm{d} x^2}{x^2 - 1} + \frac{\mathrm{d} y^2}{1 - y^2} \biggr) + 
    (x^2 - 1)(1 - y^2) \mathrm{d} \varphi^2  \biggr] \label{eq:metric_er}
\end{eqnarray}
where  
\begin{eqnarray}
  f(x,y)= \frac{x-1}{x+1} e^{-2 q P_{2}(y) Q_{2}(x)} 
  \, ,
  \label{eq:f} 
\end{eqnarray}
\begin{eqnarray}
  \gamma (x,y) = \frac{1}{2}(1+q)^{2} 
  \ln \frac{x^{2}-1}{x^{2}-y^{2}}+2 q\left(1-P_{2}(y)\right) Q_{1}(x)
  \nonumber\\
   +q^{2}
  \left(1-P_{2}(y)\right) \left[\Big(1+P_{2}(y)\Big)
  \Big(Q_{1}(x)^{2}-Q_{2}(x)^{2}\Big)\right. 
  \nonumber \\
  +\frac{1}{2} \left. \left(x^{2}-1\right)
  \Big(2 Q_{2}(x)^{2}-3 x Q_{1}(x) Q_{2}(x)+3 Q_{0}(x)  
  Q_{2}(x)-Q_{2} ^{\prime} (x)\Big)\right] 
\end{eqnarray}
The functions $P_l(y)$ and $Q_k(x)$ are the Legendre functions of first and 
second kind, respectively. Similarly to the $q$-metric, the Erez-Rosen metric 
depends on two parameters: A mass parameter $M$ and a quadrupole parameter $q$.
$M$ has the dimension of a length, whereas $q$ is dimensionless.

While the prolate spheroidal coordinates $x$ and $y$ are convenient for verifying that the 
metric satisfies the vacuum field equation, for comparing with the $q$-metric
it is advisable to transform the metric into Schwarzschild-like coordinates by
\begin{eqnarray}
    x = \frac{r}{M} - 1, \ \ \ y = \cos{\vartheta} \, .
\end{eqnarray}
In these coordinates $(t,\varphi, \vartheta, r)$  the metric was given in the 
original article by Erez and Rosen \cite{bib:er-orig}.
For a discussion of the timelike and lightlike geodesics in this vacuum 
spacetime we refer to Quevedo and Parkes \cite{bib:er-study}. We also 
mention that Castejon-Amenedo and Manko \cite{bib:er-kerr} have used the 
Erez-Rosen metric as a seed metric for constructing stationary (rotating)
solutions to Einstein's vacuum field equation which, however, will not be 
considered in the following.

For $q=0$, the Erez-Rosen metric reduces to the spherically symmetric 
Schwarz\-schild metric. Just as the $q$-metric, the Erez-Rosen metric 
is symmetric with respect to the equatorial plane $\vartheta = \pi / 2$, 
and it is asymptotically flat for any values of $(q, M)$. For $M=0$, 
irregardless of $q$, the spacetime becomes flat. If $M \neq 0$, the metric 
has a singularity at $r=2M$. In the Schwarzschild case, $q=0$, this is a 
coordinate singularity, indicating a horizon. In any other case, $q \neq 0$, 
there is a naked curvature singularity at $r=2M$. As in the case of the 
$q$-metric we may think of the Erez-Rosen metric as being valid only outside 
of a surface $\mathcal{S}$ that surrounds the naked singularity, with an 
interior matter solution matched at $\mathcal{S}$. We may then interpret 
the Erez-Rosen spacetime as describing the gravitational field around a 
deformed, compact stellar object. 

For the following calculations, the metric components $g_{tt}$ and 
$g_{\varphi \varphi}$ are of importance. In the Schwarzschild-like
coordinates, they read
\begin{eqnarray}
    g_{tt} (r , \vartheta )  = -\biggl( 1 - \frac{2M}{r}\biggr) e^{-2q P_2(\mathrm{cos} \, \vartheta ) Q_2 (r/M-1)} 
    \label{eq:gtt}
\end{eqnarray}
\begin{eqnarray}
    g_{\varphi \varphi} (r , \vartheta )  = r^2 \sin^2 \vartheta \, e^{2q P_2(\mathrm{cos} \, \vartheta ) Q_2 (r/M-1)} 
    \label{eq:gphiphi}
\end{eqnarray}
with
\begin{eqnarray}
    P_2(\cos \vartheta) = \frac{1}{2} (3 \cos^2 \vartheta - 1) \, , 
\end{eqnarray}
\begin{eqnarray}
    Q_2(r/M -1) = -\frac{1}{4} 
    \biggl(\frac{3r^2}{M^2} - \frac{6r}{M} + 2 \biggr) \ln \Big( 1 - \dfrac{2M}{r} \Big) 
    - \frac{3}{2} \biggl( \frac{r}{M} - 1 \biggr) \, .
\end{eqnarray}
The Geroch-Hansen mass monopole and quadrupole moments of the Erez-Rosen 
spacetime can be expressed as \cite{bib:mult-moments-quevedo,bib:multipole-moments}:
\begin{eqnarray}
    M_0 = M, \ \ \ M_2 = \frac{2}{15} q^3 M^3
\end{eqnarray}
We require again $M_0$ to be positive. In contrast to the $q$-metric, 
$M_0$ does not depend on the quadrupole parameter 
$q$; thus, there is no restriction on q. For any positive value of $M_0$,
the quadrupole parameter $q$, and thereby $M_2/M_0^3$, may take any value
between $- \infty$ and $\infty$. Moreover, it is important to notice
that in the Erez-Rosen spacetime a positive value of $q$ corresponds to a 
positive value of $M_2$; by contrast, in the $q$-metric (where we assumed 
$M>0$ and $q>-1$ without loss of generality) $q$ and $M_2$ had opposite signs.

The explicit form of the Kretschmann scalar of the Erez-Rosen metric will not be given here 
because of its very complicated structure, see Frutos-Alfaro et al. \cite{bib:frutosalfaro2017comparison}, 
Appendix B.
In the Schwarzschild limit $q=0$, the Kretschmann scalar reduces to 
the known expression of $\mathcal{K} = 48M^2 / r^6$, which is regular at $r=2M$.
For any other value of $q$, the Kretschmann scalar diverges at $r=2M$, 
indicating a curvature singularity. In this regard, the Erez-Rosen metric 
shows the same features as the $q$-metric.

\subsection{Lightlike and timelike circular geodesics, marginally stable and marginally bound circular orbits}

\begin{figure}[t!]
    \centering
    \includegraphics[width=\textwidth]{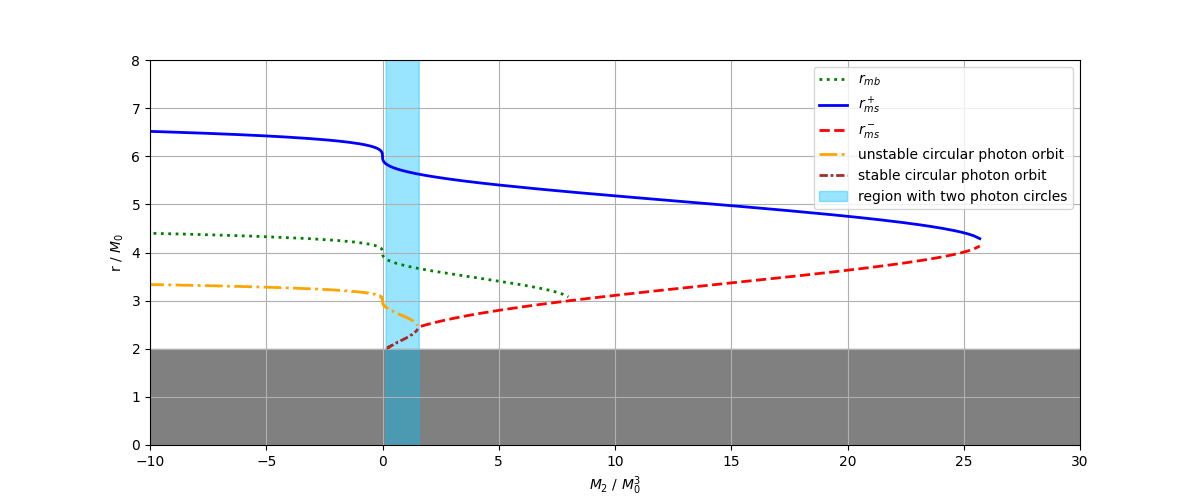}
    \caption{Circular equatorial geodesics in the Erez-Rosen spacetime, depending on the mass 
    quadrupole moment $M_2 / M_0^3$. Blue solid and red dashed line: marginally stable circular orbit; green 
    dotted line: marginally bound circular orbit; yellow and brown dash-dotted lines: positions of 
    the photon circles.}
    \label{fig:r_limits_er}
\end{figure}

For the discussion of lightlike and timelike circular geodesics in the equatorial
plane $\vartheta = \pi /2$, we insert \eqref{eq:gtt} and \eqref{eq:gphiphi} into 
\eqref{eq:Veff} to get the effective potential  
\begin{eqnarray}
    \mathcal{V} ( \varepsilon , L , r ) =  \Big( 1 - \dfrac{2M}{r} \Big) e^{q Q_2(r/M-1)} 
    \Bigg( \dfrac{L^2}{r^2} e^{q Q_2(r/m-1)} 
    + \varepsilon \Bigg)
\end{eqnarray} 
where we used that $P_2 (0) = -1/2$. 
If we insert this expression of the effective potential into \eqref{eq:circ}, 
we get the equation for circular lightlike geodesics $\varepsilon= 0$
in the equatorial plane:
\begin{eqnarray}
     r-3M + q r (r-2 M) u(r) = 0 
\end{eqnarray}
where
\begin{eqnarray}
\fl \qquad    u(r) := \dfrac{\mathrm{d}}{\mathrm{d} r} Q_2 \big(r/M-1 \big) =
     - \dfrac{3}{M}- \dfrac{M}{r \, (r-2M)} -
     \dfrac{3 \, (r-M)}{2 \, M^2} \, \mathrm{ln} \Big( 1 - \dfrac{2M}{r} \Big)
     \, .
\end{eqnarray}
There are at most two solutions of this equation which we denote $r=r_c^+$
and $r=r_c^-$; however, it depends on the value of $q$ whether they lie
outside of the singularity. For $q \leq 1$ (corresponding to $M_2/M_0^3 \lesssim 0.13$), 
there is only one photon circle, at a radius $r=r_c^+$; this photon circle is unstable. 
In the Schwarzschild limit $q = 0$, we get again the familiar value of $r_c^+ = 3M$. 
In the limiting case $q \to -\infty$, equivalent to $M_2/ M_0^3 \to - \infty$, 
we have $r_c^+ \to \infty$, whereas in the other limiting case $q = 1$, we have 
$r_c^+ \approx 2.87 M$. For $1 < q \lesssim 2.25$, corresponding to $ 0.13 \lesssim 
M_2/M_0^3 \lesssim 1.52$, there exist two photon circles at radii $r_c^+$ and $r_c^-$.
The inner one at $r_c^-$ is stable, while the outer one at $r_c^+$ is unstable. 
Beyond $q \approx 2.25$, there is no photon circle anymore. The positions of the 
photon circles are plotted in figure \ref{fig:r_limits_er}. 

\begin{figure}[t!]
    \centering
    \includegraphics[width=\textwidth]{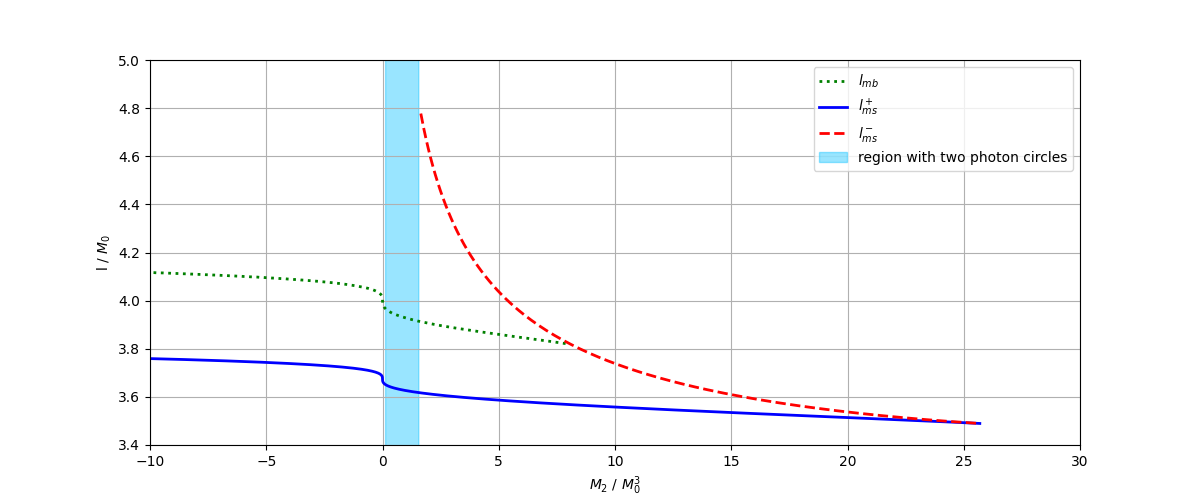}
    \caption{Specific angular momentum of the marginally stable (blue solid and red dashed lines) 
    and marginally bound orbits (green dotted line)
    in the Erez-Rosen spacetime, depending on the quadrupole moment $M_2 / M_0^3$.}
    \label{fig:l_limits_er}
\end{figure}

For timelike geodesics ($\varepsilon = 1$), the conditions \eqref{eq:circ} for circular orbits give the following expressions for the Keplerian values:
\begin{eqnarray}
    L_K ( r)^2 =   
    \dfrac{ e^{-q Q_2(r/M-1)} \, r^2 \Big( 2M+(r-2M) \, q \, r \, u(r)\Big)}{2 \Big( r-3M-(r-2M) \, q \, r \, u(r) \Big)}
    \label{eq:LKErezRosen}
\end{eqnarray}
and
\begin{eqnarray}
    E_K (r) ^2 =
    \dfrac{e^{qQ_2(r/M-1)} (r-2M)^2 \Big( 2-q \, r \, u(r) \Big)}{2 r \Big( r-3M-(r-2M) \, q \, r \, u(r) \Big)}
\end{eqnarray}
From $L_K(r)$ and $E_K(r)$, we can find the Keplerian specific angular momentum
\begin{eqnarray}
    l_K(r) ^2 = 
    \dfrac{e^{-2qQ_2(r/M-1)}  \, r^3 \Big( 2M +(r-2M) \, q \, r \, u(r) \Big)}{(r-2M)^2 \Big( 2-q \, r \, u(r) \Big)} 
    \label{eq:ksam_er}
\end{eqnarray}
and the Keplerian angular velocity
\begin{eqnarray}
    \Omega_K(r) ^2 = 
    \dfrac{e^{2qQ_2(r/M-1)} \Big( 2M+(r-2M) \, q \, r \, u(r) \Big)}{r^3 \, \Big( 2- q \, r \, u(r) \Big)} \, .
\end{eqnarray}
Following these equations, we can calculate the marginally 
stable orbits $r_{\mathrm{ms}}$ and marginally bound orbit $r=r_{\mathrm{mb}}$, with the help
of \eqref{eq:ms2} and \eqref{eq:mb}, respectively. The calculation 
can be done only numerically in both cases. We find the existence of at most two marginally 
stable orbits, at radii $r_{\mathrm{ms}}^+$ and $r_{\mathrm{ms}}^-$, and of at most
one marginally bound orbit, at radius $r_{\mathrm{mb}}$. In figure \ref{fig:r_limits_er}, 
the position of the marginally stable and marginally bound orbits are depicted, along with 
the position of the photon circles, depending on the quadrupole moment. In this figure, 
we plot on the vertical axis all radius values in units of the mass monopole moment 
$M_0$, and on the horizontal axis the quadrupole moment in terms of the asymptotically 
and invariantly defined quantity $M_2/M_0^3$. In this way, the plots for the Erez-Rosen 
spacetime are directly comparable to the plots for the $q$-metric. Note that in this 
diagram $M_2 / M_0^3$ ranges from $-\infty$ to $\infty$, whereas in the $q$-metric this 
quantity was restricted to $M_2 / M_0^3 > - 1 / 3$. 

We recall that there are two photon circles for $0.13 \lesssim M_2 / M_0^3 
\lesssim 1.52$, whereas for lower values of the quadrupole moment, there is only one, and 
for higher values, there is none. The marginally stable orbit at $r_{\mathrm{ms}}^+$ exists 
from $q= - \infty$ (corresponding to $M_2/M_0^3=- \infty$) onwards, whereas the second 
marginally stable orbit at $r_{\mathrm{ms}}^-$ comes into existence at $q \approx 2.25$ 
corresponding to $M_2/M_0^3 \approx 1.52$, i.e., at the point, where the two photon
circles merge and vanish. The two marginally stable orbits merge
at $q \approx 5.79$, or $M_2/M_0^3 \approx 25.88$. Beyond this, any 
timelike circular geodesic is stable, but note that such geodesics do not exist for all
radius values down to $r=2M$. A marginally bound orbit exists from $q = - \infty$ 
(corresponding to $M_2/M_0^3 = - \infty$) up to $q \approx 3.91$ (or $M_2 / M_0^3\approx 7.99$).

Based on these observations we find that there are three classes of Erez-Rosen spacetimes,
in close analogy to the three classes of the family of $q$-metrics, but expanded by a further
subdivision of Class II.

\begin{description}

\item[Class I]: $- \infty < q < 1$, i.e. $- \infty < M_2 /M_0^3 \lesssim 0.13$ \\
In class I, there is one unstable photon circle at radius $r_c^+$ and one marginally
stable orbit at $r_{\mathrm{ms}}^+$.  
Orbits between infinity and $r_{\mathrm{ms}}^+$ are stable, whereas orbits 
between $r_{\mathrm{ms}}^+$ and $r_c^+$ are unstable. Between $r_c^+$ and $2M$,
there are no timelike circular geodesics. This class, which contains the 
Schwarzschild case with $q=0$, is completely analogous to Class I of the 
$q$-metric case.

\item[Class IIa]: $1 < q \lesssim 2.25$, i.e., $0.13 \lesssim M_2 /M_0^3 \lesssim 1.52$ \\
In this class, we have one marginally stable orbit at radius $r_{\mathrm{ms}}^+$
and two photon circles, an unstable one at radius $r_c^+$ and a stable
one at radius $r_c^-$. We have stable circular timelike geodesics above 
$r_{\mathrm{ms}}^+$, unstable ones between $r_{\mathrm{ms}}^+$ and $r_c^+$, 
no such orbits between $r_c^+$ and $r_c^-$, and again stable ones below
$r_c^-$; although, they may not necessarily exist all the way down to $r=2M$.
The presence of two stable regions that are separated from each
other is similar to $q$-metrics of Class II. However, the occurrence of 
a region with no timelike orbits between the two stable regions is a key difference
in comparison to the $q$-metric.

\item[Class IIb]: $2.25 \lesssim  q \lesssim 4.8$, i.e., $1.52 \lesssim M_2 /M_0^3 \lesssim 25.8$ \\
In this class, we have two marginally stable orbits at radii $r_{\mathrm{ms}}^+$
and $r_{\mathrm{ms}}^-$. Timelike circular geodesics are stable above $r_{\mathrm{ms}}^+$ and below
$r_{\mathrm{ms}}^-$, as far as they exist, and unstable between these values. This situation is 
completely analogous to Class II of $q$-metrics. 

\item[Class III]: $4.8 \lesssim q < \infty$, i.e., $25.8 \lesssim M_2/M_0^3 < \infty$ \\
In class III, there are no marginally stable circular orbits. All circular 
timelike geodesics are stable; however, circular timelike geodesics do
not exist all the way down to $r=2M$. Again, this is analogous to Class
III of $q$-metrics.

\end{description}

\begin{figure}[t!]
    \centering
    \includegraphics[width=\textwidth]{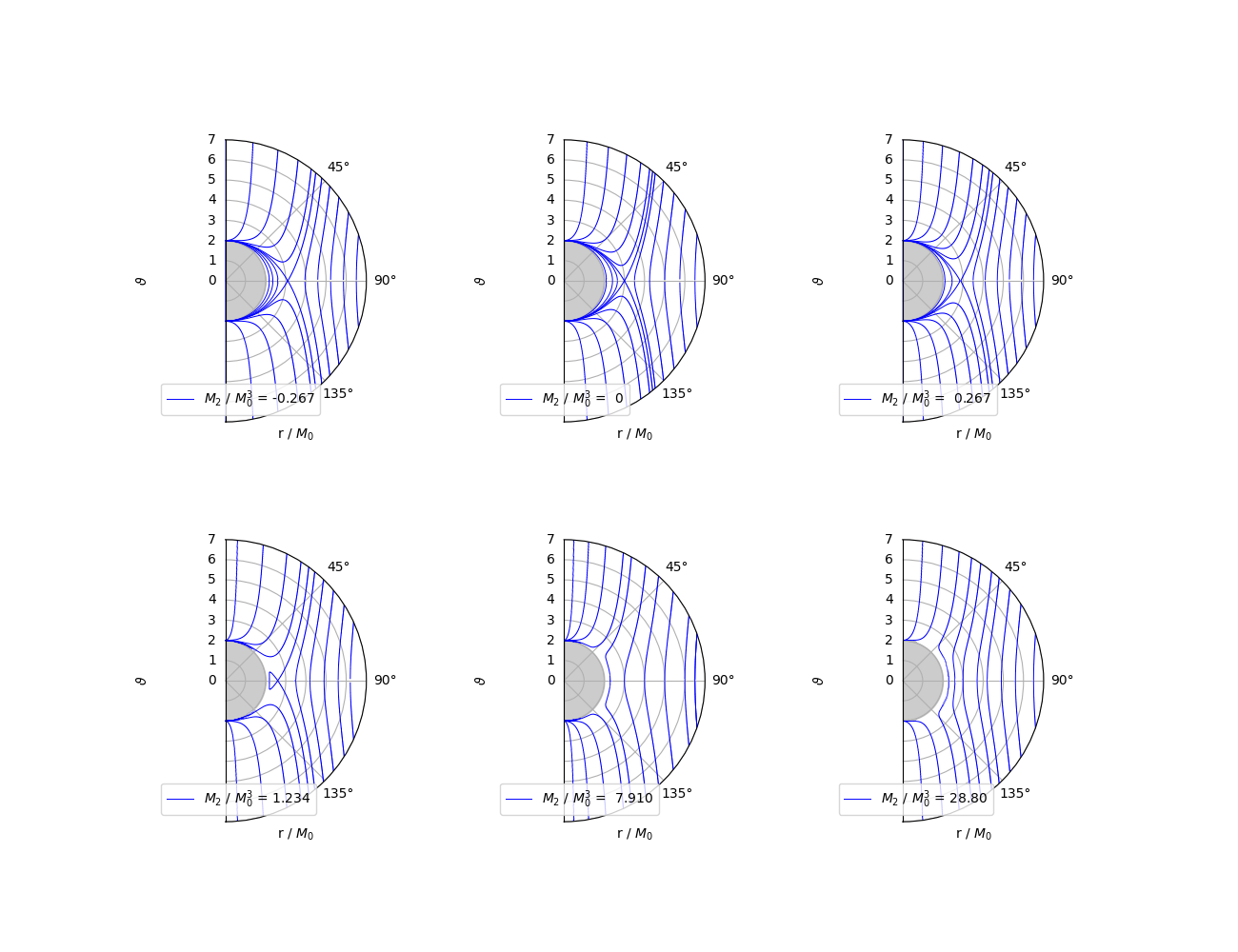}
    \caption{Von-Zeipel cylinders for different mass quadrupole moments in the Erez-Rosen spacetime. Upper row left: Class I spacetime with a negative value of $M_2 / M_0^3$, corresponding to an oblate geometry; upper row middle: Schwarzschild spacetime; upper row right: Class I spacetime with a positive value of $M_2 / M_0^3$, corresponding to a prolate geometry; lower row left: Class IIa spacetime; lower row middle: Class IIb spacetime; lower row right: Class III spacetime.}
    \label{fig:vzc_er}
\end{figure} 

The specific angular momentum of the marginally stable and marginally bound orbits can be 
found by inserting the radius values $r_{\mathrm{ms}}^{\pm}$ and $r_{\mathrm{mb}}$, respectively, 
into the expression for the Keplerian specific angular momentum in Eq. \eqref{eq:ksam_er}. 
The resulting values of $l_{\mathrm{ms}}^{\pm}$ and $l_{\mathrm{mb}}$ are plotted in 
figure \ref{fig:l_limits_er}, depending on $M_2 / M_0^3$. 

\subsection{Von-Zeipel cylinders}

In this section, the von-Zeipel cylinders are calculated in the 
Erez-Rosen spacetime. They give an insight into the behaviour of 
fluids in circular motion in this spacetime.
Recall that the von Zeipel cylinders are defined as the surfaces 
$\mathcal{R} = \textrm{const.}$, with $\mathcal{R}$ given by 
Eq. \eqref{eq:vzc_general}, and that they are independent of 
$l$ (or $\Omega$). In the Erez-Rosen spacetime, $\mathcal{R}$ 
reduces to
\begin{eqnarray}
    \mathcal{R}(r, \vartheta )^2 = e^{4q P_2(\mathrm{cos} \, \vartheta ) Q_2 (r/M-1)} 
    \, \dfrac{r^2}{\big( 1 - \frac{2M}{r}\big)} \ .
    \end{eqnarray}
Plots of the von Zeipel cylinders can be found in figure \ref{fig:vzc_er}.

As in any other asymptotically flat spacetime, the von Zeipel cylinders approach
flat cylinders far away from the vertical axis.
At the position of an unstable photon circle, the corresponding cylinder has a self-crossing, as 
is the case e.g. in the Schwarzschild spacetime at $(r = 3M_0,\vartheta = \pi /2)$. 
For negative values of $M_2 / M_0^3$, the radius coordinate of the unstable photon circle
is bigger than $3M_0$; for positive values of $M_2 / M_0^3$, it is smaller.
Beyond $M_2 / M_0^3 \approx 1.56$, where there is no photon circle anymore, 
no von Zeipel cylinder has a self-crossing.

Note that, in Erez-Rosen spacetimes of Class IIa, there are von Zeipel ``cylinders'' 
that actually have the topology of a torus. Such ``von Zeipel tori'' occur near 
a stable circular light ray, see the picture on the left in the lower row of
figure \ref{fig:vzc_er}. They have been observed even in spherically symmetric 
and static spacetimes before, e.g. in interior Schwarzschild solutions, see 
Abramowicz et al. \cite{bib:AbramowiczEtAl1993}.

\subsection{Effective potential}

\begin{figure}[t!]
    \centering
    \includegraphics[width=1.1\textwidth]{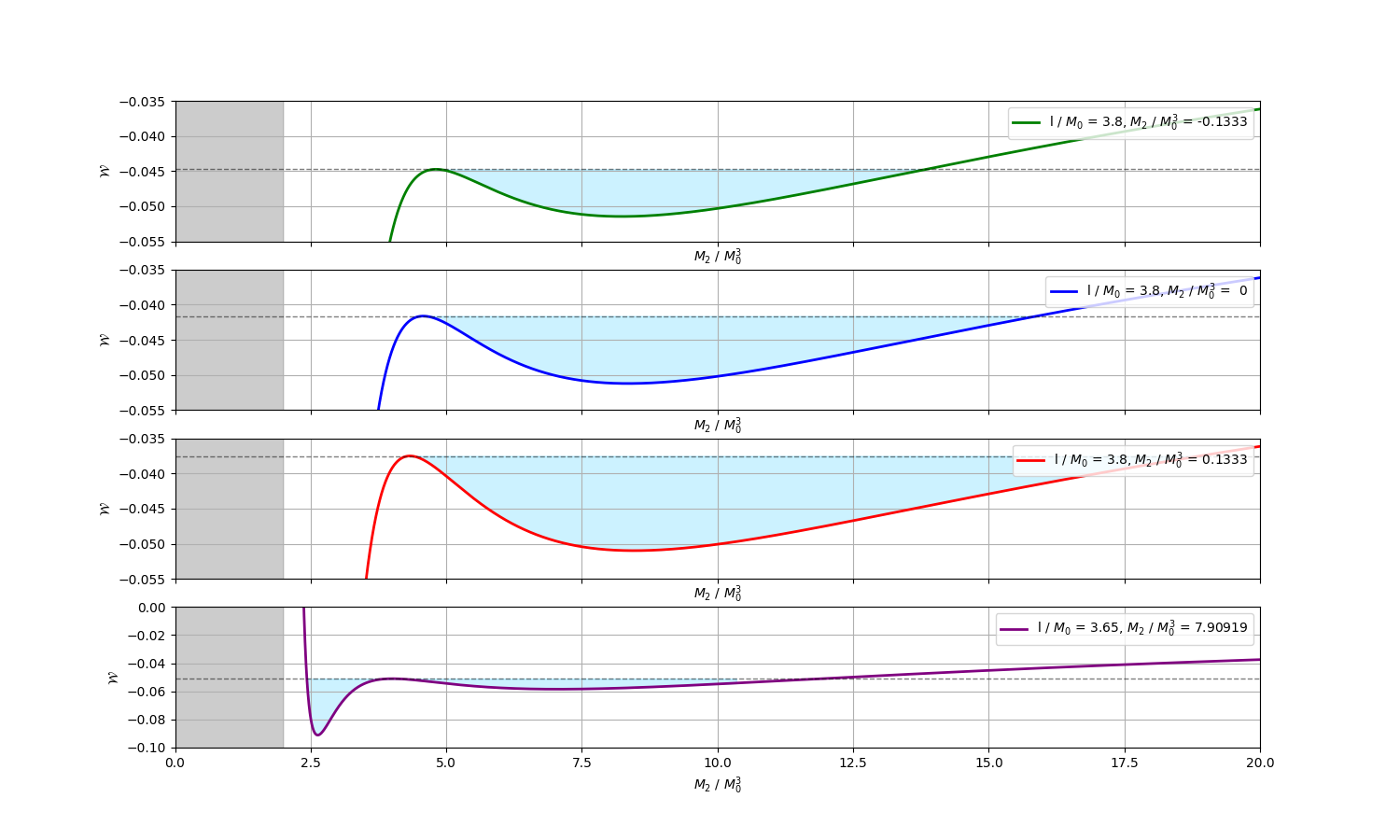}
    \caption{Effective potential
    in the equatorial plane in Erez-Rosen spacetime, depending on $M_2 / M_0^3$. 
    The grey shaded area indicates the region bounded by the singularity at $r=2M_0$, the 
    lighter blue-shaded area is the region occupied by a torus filling its
    Roche lobe completely. Upper graph: Class I
    spacetime with a negative value of $M_2 / M_0^3$, corresponding to an oblate geometry; 
    upper-middle graph: Schwarzschild spacetime; 
    lower-middle graph: Class I spacetime with a positive value of $M_2 / M_0^3$, corresponding 
    to a prolate geometry; lower graph: Class IIb spacetime, note that effective potential approaches $+\infty$
    for $r \to r_{\mathrm{sing}}$.}
    \label{fig:eff_pot_er}
\end{figure}

\begin{figure}[t!]
    \centering
    \includegraphics[width=\textwidth]{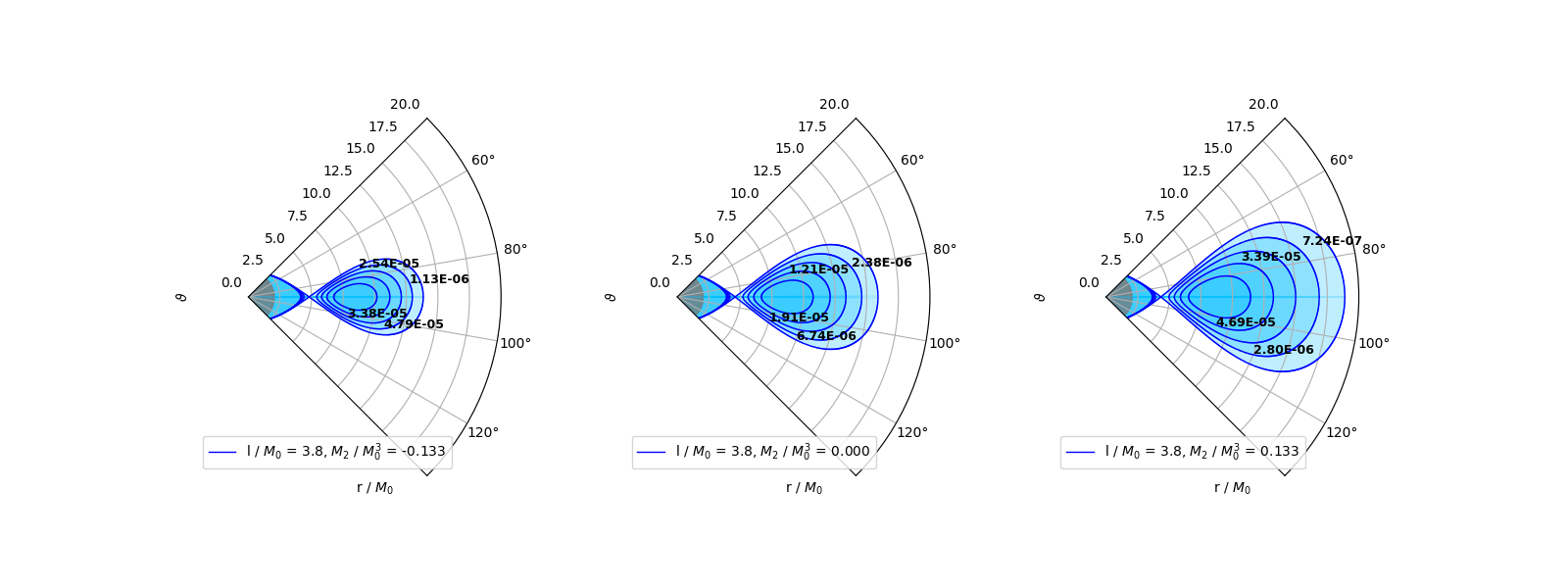}
    \caption{Polar dependency of the effective potential in the Erez-Rosen metric; black numbers represent the density in units of $M^{-2}$ where we assumed, again, a polytropic equation of state with $K=0.2$ and 
    $\Gamma = 5/3$. For each torus we compare the density $\rho$ with the Kretschmann scalar $\mathcal{K}$ 
    at the centre to demonstrate that the self-gravity of the fluid is negligible.
Left: Class I spacetime with a positive value of $M_2 / M_0^3$, corresponding to a prolate geometry,
$\rho(r_{\mathrm{cen}})^2 / \mathcal{K}(r_{\mathrm{cen}}) = 5,45 \times 10^{-5}$; middle:
Schwarzschild spacetime, $\rho(r_{\mathrm{cen}})^2 /\mathcal{K}(r_{\mathrm{cen}}) = 1.18 \times 10^{-6}$ ; right: Class I spacetime with a negative values of
$M_2 / M_0^3$, corresponding to an oblate geometry, $\rho(r_{\mathrm{cen}})^2 / \mathcal{K}(r_{\mathrm{cen}}) = 2.39 \times 10^{-5}$.}
    \label{fig:eff_polar_er}
\end{figure}

\begin{figure}[ht!]
    \centering
    \includegraphics[width=1.1\textwidth]{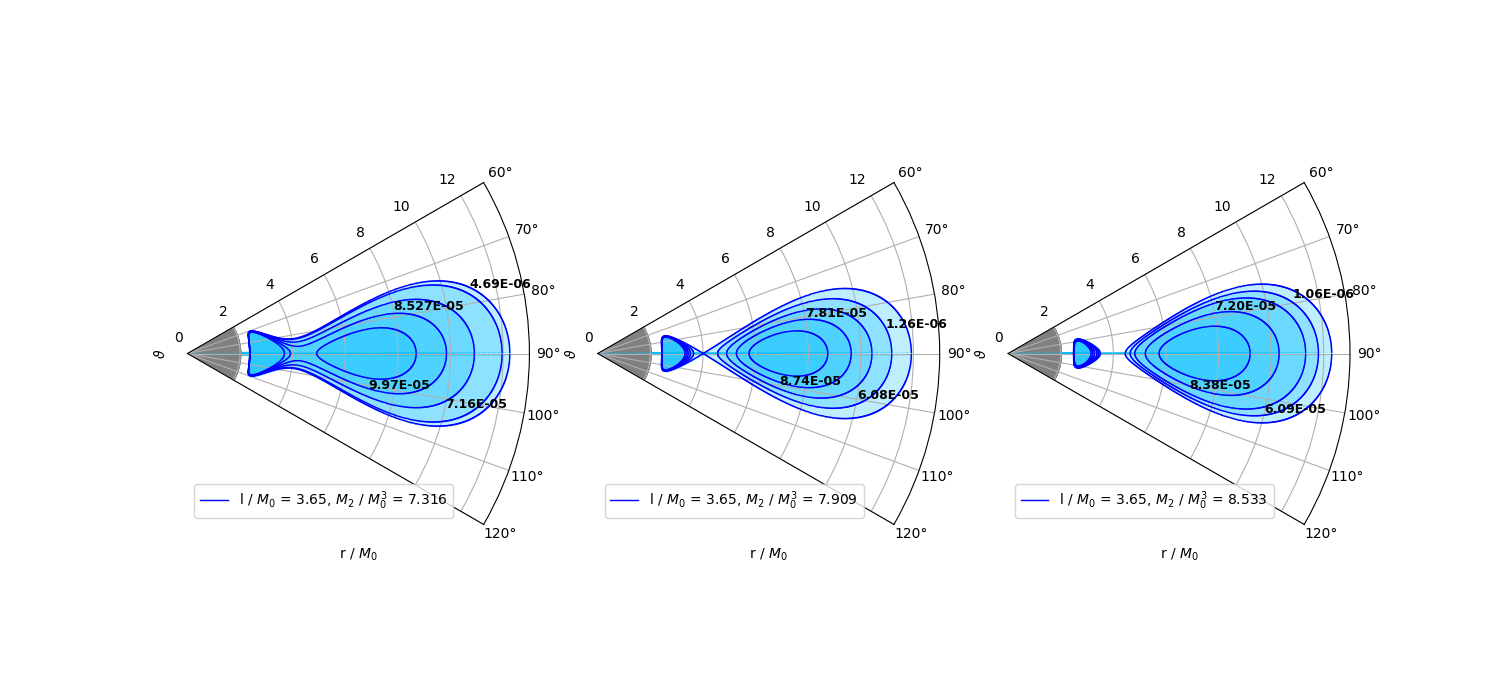}
    \caption{Polar dependency of the effective potential in an Erez-Rosen metric of Class II, for connected 
    double tori (left panel), touching double tori (middle panel) and spatially separated tori (right panel). 
    The black numbers represent the density in units of $M_0^{-2}$ where we assumed again a 
    polytropic equation of state with $K=0.2$ and $\Gamma = 5/3$: 
    left: $\rho(r_{\mathrm{cen}})^2 / \mathcal{K}(r_{\mathrm{cen}}) = 2.66 \times 10^{-5}$;
    middle: $\rho(r_{\mathrm{cen}})^2 / \mathcal{K}(r_{\mathrm{cen}}) = 2.33 \times 10^{-5}$;
    right: $\rho(r_{\mathrm{cen}})^2 / \mathcal{K}(r_{\mathrm{cen}}) = 2.04 \times 10^{-5}$.}
    \label{fig:fish_er}
\end{figure}

To construct Polish doughnuts in the Erez-Rosen spacetime, 
we calculate the effective potential (see Eq. \eqref{eq:eff_pot_general}):
\begin{eqnarray}
    \mathcal{W} = \frac{1}{2} \ln{\biggl[\frac{r^2 (r - 2M) \mathrm{sin} ^2 \vartheta \, e^{-2q P_2(\mathrm{cos} \, \vartheta ) Q_2 (r/M-1)}}{r^3 \mathrm{sin} ^2 \vartheta - l^2 (r - 2M) e^{-4q P_2(\mathrm{cos} \, \vartheta ) Q_2 (r/M-1)} }\biggr]} \, .
    \label{eq:eff_pot_er}
\end{eqnarray}
As before in the $q$-metric, we plot this potential in the equatorial plane 
for four different pairs of values ($l / M_0, \, M_2 / M_0^3$) in figure \ref{fig:eff_pot_er}, 
and we plot correspondingly constructed Polish doughnuts in figures \ref{fig:eff_polar_er} and 
\ref{fig:fish_er}. We note that, again, the effective potential satisfies condition
\eqref{eq:Wtheta} and that, for any values of $q$ and $l$, it approaches 0 from below for 
$r \to \infty$. 

For spacetimes of Class I, we have, in complete analogy to the $q$-metric case, one
marginally stable orbit at radius $r_{\mathrm{ms}}^+$, a marginally bound orbit at 
radius $r_{\mathrm{mb}}$ and one unstable photon orbit at radius $r_c^+$, where 
$2M < r_c^+< r_{\mathrm{mb}}<r_{\mathrm{ms}}$. 
The potential $\mathcal{W}(l,r,\pi/2)$ approaches $- \infty$ for $r \to 2M$. For 
$l^2 > l_{\mathrm{ms}}^+$, it has a minimum at $r_{\mathrm{cen}}$, 
and maximum at $r_{\mathrm{cusp}}$. Therefore, by choosing 
$l^2 > l_{\mathrm{ms}}^+$, we can construct the same
type of doughnuts as for Class I in the $q$-metric case, 
see the first three panels in figure \ref{fig:eff_pot_er} and figure \ref{fig:eff_polar_er}.
In all cases, we have chosen $l^2< l_{\mathrm{mb}}$, guaranteeing that the Roche
lobe features a cusp, and we have chosen $\mathcal{W}_{\mathrm{in}}$ in such a way that the
fluid fills its Roche lobe completely. In this way we get Polish doughnuts that
are quite similar to the tori in the Schwarzschild spacetime. Of course, the shape of the 
Roche lobe is influenced by the quadrupole moment, but qualitatively there is no 
difference to the Schwarzschild case.

In spacetimes of Class II, the doughnuts are similar to those in the $q$-metric
case of the corresponding class. Recall that here Class II is subdivided into
Class IIa and Class IIb. In Class IIa, there is one marginally stable
orbit at radius $r_{\mathrm{ms}}^+$, a marginally bound orbit at radius
$r_{\mathrm{mb}}$ and two photon circles at radii $r_c^{\pm}$, where
$2M < r_c^-<r_c^+<r_{\mathrm{mb}} < r_{\mathrm{ms}}^+<\infty$; in the 
Class IIb, there is one marginally stable orbit at radius $r_{\mathrm{ms}}^+$,
possibly a marginally bound orbit at radius $r_{\mathrm{mb}}$, and another
marginally stable orbit at radius $r_{\mathrm{ms}}^-$, where $2M<r_{\mathrm{ms}}^-
(<r_{\mathrm{mb}})< r_{\mathrm{ms}}^+<\infty$. In either case, there are two 
separated stable regions, and this is the reason why Polish doughnuts in 
Class IIa spacetimes are qualitatively similar to the ones in Class IIb spacetimes.
By choosing $l^2 > (l_{\mathrm{ms}}^+)^2$, we can construct the
same kind of double torus as in the $q$-metric case, see the lower panel 
of figure \ref{fig:eff_pot_er} and figure \ref{fig:fish_er}. As the latter
figure demonstrates, the cross-section of the Roche lobes of such double 
tori indicate the same fish-like shape as in the $q$-metric case. Again, we 
emphasise that the ``fish-tail'' is isolated from the singularity at 
$r=2M$. As long as the marginally bound orbit exists, we have to choose 
$l^2 < l_{\mathrm{mb}}$ if we want to have two finite Roche lobes 
that meet in touching cusps. If the marginally bound orbit does not exist anymore, 
any value $l^2 > l_{\mathrm{ms}}^+$ will do. As in the case of the $q$-metric,
we can choose $\mathcal{W}_{\mathrm{in}}$ so big that the double torus 
becomes a single torus with two centres, see the picture on the left in
figure \ref{fig:fish_er}.

Finally, in Erez-Rosen spacetimes of Class III, all timelike circular 
geodesics are stable. Therefore, we have exactly the same
type of Polish doughnuts without a cusp as in the corresponding $q$-metric
case. 

Note that, again, we have assumed that the spacetime geometry is given by the unperturbed
Erez-Rosen metric, neglecting the self-gravity of the torus. 
This is justified only as long as the density of the fluid is small enough. For the examples 
shown in figure \ref{fig:eff_polar_er} and \ref{fig:fish_er}, we demonstrate that indeed
at the centre of the torus the density squared is much smaller
than the Kretschmann scalar.

\newpage

\section{Conclusions}

In this paper, we discussed geometrically thick accretion tori in two quadrupolar spacetimes, 
the $q$-metric and the Erez-Rosen spacetime.
It was our main goal to find out if there are new qualitative features that distinguish these tori from
analogously constructed ones in the Schwarzschild spacetime. We have seen that, both in the $q$-metric and
in the Erez-Rosen spacetime, there are three classes to be distinguished, depending on the value of the
Geroch-Hansen quadrupole moment $M_2$ (choosing the ADM mass $M_0$ as the unit of length). 
Class I comprises all spacetimes with (allowed) negative values of 
$M_2$ and also those with positive $M_2$ up to a certain value $M_2^{(1)}$. The tori in this class are
similar to the tori in the Schwarzschild spacetime, which is contained as a special case
in Class I. While the precise shape of a torus and the position of its outer and inner edges do
depend on the quadrupole moment, there is no influence on the qualitative features. Then there 
is a Class III which comprises all spacetimes with positive quadrupole moment bigger than 
a certain value $M_2^{(2)}$. The major difference compared to Class I is in the fact that the 
tori \emph{cannot} have a cusp, i.e., they cannot be at the verge of accretion, 
whereas in Class I the torus has a cusp if the parameters are chosen appropriately.
Interesting new features arise in Class II, i.e., in spacetimes with a 
positive quadrupole moment between $M_2^{(1)}$ and $M_2^{(2)}$. 
There it is possible to construct \emph{double tori}, both of finite size and bounded away
from the naked singularity and from infinity. The meridional cross-section of such a double torus
has the shape of a fish if the two Roche lobes are filled completely, with the outer torus corresponding
to the body of the fish and the inner one to the fish-tail. As the inner Roche lobe has a cusp on
its outer side and the outer Roche lobe has a cusp on its inner side, the fluid does not ``spill over''
if it extends beyond the Roche lobes; rather, it forms a single torus with two centres. 
These double tori, and also the single tori with two centres, are new features which do not occur
in the Schwarzschild spacetime. However, perfect-fluid configurations in the form of double tori 
have also been found in the Kerr spacetime, for certain values of the parameters, by Pugliese et al. 
\cite{PuglieseMontani2015,PuglieseStuchlik2017}. Moreover, it is known that \emph{charged} fluid 
configurations can form such double tori on the Schwarzschild spacetime in the presence of 
a magnetic field, see Trova et al. \cite{trova2020}. In particular in the latter case, the 
double tori are similarly shaped as the ones we have found here. So, if we ever observe 
such double tori we will have to investigate if there are any indications for 
a significant net charge of the fluid and for the presence of a magnetic field. 
If there are no such indications, then it would be justified to conjecture that the 
central object is a black hole impostor with a non-zero quadrupole moment.

\section*{Acknowledgements}

We wish to thank Dennis Philipp for an important comment. Moreover, we 
gratefully acknowledge support from Deutsche Forschungsgemeinschaft within 
the Research Training Group 1620 ``Models of Gravity”.

\section*{References}

\bibliographystyle{iopart-num}

\end{document}